\documentclass[
reprint,
 amsmath,amssymb,
 iop,
]{revtex4-2}
\usepackage{braket}
\usepackage{algorithm2e}
\usepackage{float}
\usepackage{graphicx}
\usepackage{color}
\usepackage{amsmath}
\usepackage{amsfonts}
\usepackage{amssymb}
\usepackage{array}
\usepackage{framed}
\usepackage{soul}
\usepackage[normalem]{ulem}
\providecommand{\U}[1]{\protect\rule{.1in}{.1in}}
\usepackage{hyperref}
\hypersetup{
    colorlinks=true,
    linkcolor=blue,
    filecolor=magenta,      
    urlcolor=cyan,
}

\begin{document}

\title{Unraveling the significance of Raman modes, Grüneisen parameters and phonon lifetimes in the hexagonal allotropes of Silicon and Germanium compounds}
\author{Lekshmi S. M.}
\affiliation{Department of Physics, Birla Institute of Technology and Science Pilani Hyderabad Campus, Jawaharnagar, Hyderabad, Telangana 500078, India.}
\author{Upasana Agrawal} 
\affiliation{Department of Mathematics, Birla Institute of Technology and Science Pilani Hyderabad Campus, Jawaharnagar, Hyderabad, Telangana 500078, India.}

\author{Akarsh Jain}
\affiliation{Department of Computer Science and Engineering, Birla Institute of Technology and Science Pilani Hyderabad Campus, Jawaharnagar, Hyderabad, Telangana 500078, India.}

\author{Siddharth Sastri}
\affiliation{Department of Computer Science and Engineering, Birla Institute of Technology and Science Pilani Hyderabad Campus, Jawaharnagar, Hyderabad, Telangana 500078, India.}

\author{Suvadip Das}
\thanks{corresponding author}
\email{suvadip.das@hyderabad.bits-pilani.ac.in}
\affiliation{Department of Physics, Birla Institute of Technology and Science Pilani Hyderabad, Jawaharnagar, Hyderabad, Telangana 500078, India.}

\date{\today}
\begin{abstract}
 Advancement in quantum information and quantum technologies has ushered in a new era of technological revolution in large scale atomistic simulation and efficient system on a chip device fabrication. This has led to innovative ways of harnessing rigorous search algorithms for functional quantum materials and steered scientists to dig deeper into the world of quantum phenomenon and applications. In this work, we delineate the advanced electronic structure and vibrational properties utilizing the popular meta-GGA functionals, spectral signatures of the Raman active phonon modes, explored their average mean free paths, and whether they conserve helicity, by leveraging first principles density functional theory and density functional perturbation theory. A systematic analysis of the role of phonon lifetimes, consequences of phonon-phonon and three phonon scattering rates and phonon linewidths have been presented. Further, a study of the the frequency and temperature dependent Grüneisen parameter has been employed in conjecture with the temperature dependent thermal expansion and thermal conductivity to portray the effect of anharmonicity in the phonon spectra of these two materials. Finally, we provide strategies for tuning the properties of these materials in an effort to improve their efficacy for advanced thermoelectric, photovoltaic and optoelectronic device applications.
\end{abstract}

\maketitle

\section{Introduction}
With the advancement of exascale computing facilities worldwide, quantum computing and large scale atomic simulations have gained significant impetus. The search for high performance quantum materials with improved efficiency have motivated scientists to delve into the realm of two-dimensional materials relevant for quantum technologies. Cubic silicon, has been the backbone of the semiconducting industry over the last few decades. Its abundance and cost-effectiveness has led to technological applications in power circuits, integrated chips, photovoltaics and light emitting diodes (LEDs) and transistors \cite{Ballif2022}, \cite{Xiao2023}. The immense popularity of cubic silicon in the semiconducting industry has naturally extrapolated its usefulness to fabricating quantum dots, quantum computing devices and quantum sensors in recent years.\\

Despite the popularity of silicon, drawbacks such as its brittleness, limited compatibility for interfaces, increased power consumption and heat generation, have stunted the progress of large scale silicon-based electronics \cite{Ballif2022}. Over the last couple of decades, a significant scientific effort has been devoted to the discovery of novel quantum materials, that would replace cubic silicon in technological applications.  This 
encompasses a broad range of materials from 2D halide perovskites, transition metal dichalcogenides (TMDs), 
Group IV – VI compounds and unveils unique unexplored aspects of novel quantum phenomenon for efficient energy applications \cite{Wu2021}. This has motivated the exploration of novel quantum phenomenon have led to the establishment of 
benchmarking formalisms comprising of various levels of theory ranging from molecular dynamics, density 
functional theory, density functional perturbation theory and dynamical mean field theory to many-body techniques such as quasiparticle GW approximation specifically for quantum mechanical systems with large scale atomic 
arrangement.\\

Inspite of a thorough scientific investigation into a plethora of quantum materials, there has been a lot less emphasis on the prototypical polymorphs of silicon and germanium. Silicon and germanium, in their natural diamond cubic form (space group $Fd\overline{3}m$
) are already successful and have been extensively utilized in quantum applications owing to their exceptional electronic and thermal properties \cite{si_ge}. Recently, alternate allotropes of the cubic phase, such as the $\beta$ tin, simple hexagonal, body centered and Lonsdaleite or hexagonal diamond phases have been explored for improved functionality, stability and optoelectronic properties. With the advent of data-assisted materials discovery and predictions harnessing machine learning algorithms, a series of new potential phases of silicon has been revealed as potential optoelectronic, thermoelectric and photovoltaic materials \cite{Wang2024}. The predicted allotropes range from a variety of space groups such as Im3$-$m$\,$ (229), $C2/c\,$(15), $I4/mcm\,$ (140), $I4/mmm\,$ (139), $P21/m\,$ (11), and $P4/mbm\,$ (127)  \cite{wippermann2016novel} with the possibility of hosting direct band gaps, a potential criterion for efficient photovoltaic applications. Interestingly, first principle calculations suggest the band gap of these phases of silicon lie between 0.65 t0 1.47 eV, making them ideal candidates for the generation of multiple electron-hole pairs with improved absorption and reduced recombination rates \cite{Wei2019}.\\

Among these polymorphs, particularly the hexagonal 2-H polytype (also referred to as lonsdaleite) has attracted considerable attention due to their smaller band gaps and potential optoelectronic applications. Despite hosting an indirect band gap, hexagonal silicon shows possibility of conversion to a direct band gap semiconductor under application of large bi-axial strain \cite{Rdl2015}. Hexagonal silicon in the lonsdaleite phase crystallizes in the $P63/mmc$ space group with $D_{6h}^4$ symmetry and exhibits a multitude of promising properties that differentiate it from the other allotropes of silicon \cite{hex_si_realized}.\\

Another material, hexagonal 2-H germanium, also occuring in the lonsdaleite phase, shows potential for optoelectronic applications due to the possibility of generation of multiple electron-hole pairs leading to improved optical absorption \cite{rodl2019hexge}. Unlike hexagonal silicon, reported to host an indirect band gap, 2-H lonsdaleite germanium has been predicted to exhibit a small direct band gap contrary to that of silicon \cite{De2014}. As opposed to the allotropes of silicon, there are few literatures till date that  detail the electronic, vibrational and optical properties of hexagonal polymorph of germanium. Alloys of Si/Ge have shown immense prospect in industrial applications due to the tunability of the band gaps with doping and strain, high electron mobility and efficient thermoelectric capabilities \cite{Borlido2022}, \cite{Fan2018}, \cite{hexge_expt_bg}, \cite{thermal_si_ge_hex}. Recently, SiGe heterostructures and quantum dots have been observed to host spin qubits with high coherence, thereby resulting in a cascade of studies focusing on these materials and their polymorphs to address their indispensable utilization in quantum computing and information storage devices \cite{Thayil2025}. Hence, a comprehensive study of the electronic and vibrational properties of the hexagonal polymorphs of silicon and germanium is quite timely.\\

In spite of the increasing popularity of the silicon and germanium compounds, earlier studies suggest that some of the polymorphs are either unstable, metastable or stable only under high pressure conditions. Recent studies utilizing machine learning algorithms and mapping of energetics have suggested the possibility of occurrence of stable polymorphs even at ambient conditions. Albeit earlier claims, in the last few years, multiple experimental investigations have suggested the stable growth of the 2-H Si and Ge in the form of heterostructures as well as on top of nanowires \cite{Dushaq2019}, \cite{Tizei2025}. Hence, the occurence of stable polymorphs of hexagonal silicon and germanium are still much of a possibility and warrants further scientific investigation.\\ 

The material of interest in our study, 2-H lonsdaleite silicon has been synthesized experimentally on hexagonal gallium phosphide using crystal structure transfer method \cite{hex_si_realized}. The hexagonal 2-H polymorph of germanium was first synthesized at low pressures employing UV laser ablation technique \cite{HExGeexpt}. The nano-indentation of germanium grown on silicon substrate, has also seen to result in a phase transition to the hexagonal phase at room temperature under pressure of 4.9 to 8.1 GPa \cite{Dushaq2019}. Compared to the 2C cubic phase, the reduced crystal symmetry has been suggested to result in additional optical phonon modes which facilitate increased scattering and contribute to a significant reduction in the thermal conductivity of hexagonal silicon compared to cubic silicon \cite{raya2017thermal}. Thermal conductivity measurements decide whether a material would indeed be a promising candidate for thermoelectric applications.\\

The innumerable possibilities of the allotropes of silicon and germanium in potential device applications, particularly in the 2-H phase, warrants a detailed quantitative analysis of the electronic, vibrational and optical properties of these materials. Both lonsdaleite silicon and germanium are potential optoelectronic materials with relative smaller predicted band gaps with potential generation of multiple electron-hole pairs, band topologies resulting in large carrier concentrations and mobility. Further, unlike hexagonal silicon, 2-H germanium is a direct band gap semiconductor with a smaller gap compared to silicon, arising primarily due the folding of the L point of the cubic germanium into the $\Gamma$ point \cite{rodl2019hexge}. Apart from potential industrial applications, hexagonal polymorphs of silicon and germanium offer a suitable platform for the benchmarking of various levels of first-principles theoretical frameworks.\\

Inspite of the importance of the properties of these materials and experimental investigation, there exists only a handful of first principles calculations of the electronic, and structural properties of these materials. The difficulty in predicting accurate band gaps stems from the underestimation of the binding energy of 3d electrons by conventional exchange-correlation functionals. The repulsion between the electronic states and screening in this material are often not accurately accounted for leading the valence band maximum of the p orbital to be of higher energy, hence underestimating the band gap. For hexagonal polymorph of germanium, the popular HSE06 hybrid functional scheme is found to result in a band gap of 0.286 eV while the modified Becke-Johnson (mBJLDA) LDA functional of meta-GGA type estimates the band gap to be 0.298 eV\cite{rodl2019hexge}. Another study has obtained a bandgap of 0.32 eV using one of the flavours of Hybrid functionals, the HSE06 functional\cite{hexge_hse06}.  Photoluminescence spectroscopy measurements performed on experimentally grown samples reveal the optical band gap at low temperatures to be 0.353 eV \cite{hexge_expt_bg}. Further, accurate evaluation of the band gap in this material is found to be highly sensitive to applied strain and growth conditions of  the sample.\\

Along with the estimation of band gaps, ascertaining the stability of these compounds and their interfaces are important for device applications. The evaluation of the thermal conductivity requires a detailed analysis of the phonon spectrum in these materials. Both silicon and germanium in the hexagonal phase have been reported to exhibit anisotropic thermal conductivities at room temperature \cite{Borlido2022}. Recent studies reveal the possibility of significant contributions from four phonon scatterings to the anisotropic thermal conductivities \cite{thermal_si_ge_hex}. Most first principles and density functional perturbation theory calculations on these materials were performed at absolute zero temperature under the harmonic approximation. These results, when extrapolated to non-zero temperatures, lead to the so called quasi-harmonic approximation (QHA), where the phonon frequencies are considered to change with the volume of the lattice structure. However, the above approach is not always accurate at very high temperatures and in the vicinity of phase transitions. Further the QHA approximation neglects the effect of phonon-phonon anharmonicity, which necessitates the need for the incorporate of explicit temperature dependence in the phonon calculations \cite{Gonze2020}.
 


Despite significant efforts to characterize the electronic properties of hexagonal silicon and germanium, there remain contradictions and caveats in the understanding of their electronic structure, phonon frequencies, dynamics, raman spectroscopy and thermal transport properties. In particular, the role of anharmonic phonon-phonon interactions and their impact on thermal conductivity in these low-symmetry materials have not been systematically studied from first principles. Another unanswered question lies in the effect of screening and level repulsions in the various recent implementations of the popular exchange correlational schemes. Understanding these mechanisms is essential to assess their potential for photovoltaic and thermoelectric devices, and would lead to progress in the theoretical modeling of thermal transport in low-dimensional and low-symmetry materials.\\

In an attempt to address the lack of rigorous quantitative analysis and controversies regarding the results in photovoltaic and thermoelectric properties of the silicon and germanium polymorphs, we employed ab initio methods for a detailed  evaluation of the electronic structure utilizing improved exchange correlation functionals, mode decomposed grüneisen parameters, phonon dispersion, and signatures of raman active modes in theoretical spectroscopy. The manuscript is arranged as follows: Section I provides an introduction and motivation of our study, Section II delineates the theoretical background and computational methodologies, Section III illustrates the results, research findings and discussions whereas Section IV summarizes the conclusions from our detailed scientific investigation.

\section{Theoretical and Computational Methods}

In this section of the manuscript, we delineate the theoretical concepts and computational formalism utilized in our study of the silicon and germanium polymorphs of interest. Over the years, a plethora of computational techniques have been applied to decipher the fingerprints of quantum phenomenon in novel functional materials. The popular  formalisms include but are not limited to wavefunction-based theoretical approaches, density functional theory, dynamical mean field theory, auxilliary field-based monte carlo method and green's function approach. Except for a handful of cases involving strongly correlated electrons, density functional theory has been the framework of choice for theoretical characterization of materials properties as well as validation of experimental probes such as X-ray photoelectron spectroscopy (XPS), angle-resolved photoemission spectroscopy (ARPES),  and vibrational Raman spectroscopy till date.\\

In the remainder of the section, we aim to provide the theoretical details of our first-principles formalism including density functional theory (DFT) and density functional perturbation thery (DFPT) for the quantitative analysis of the electronic structure, nature of the band gap, raman modes and stability of the compounds. We introduce concepts and formalisms for computing the effects of anharmonicity in our phonon calculations, phonon linewidths and lifetimes. Finally, we delineate the computational scheme for the theoretical investigation of raman-active modes in the phonon dispersion, temperature dependent grüneisen parameters, transport and thermal properties for device applications in thermoelectrics and photovoltaics.

\subsection{Density functional theory and Density functional perturbation theory}

In this study, we employ first principles methodologies for examining the electronic spectrum and structural properties within the framework of the density functional theory as implemented in the Quantum espresso package.\cite{QE-2009} Albeit the reliability and applicability of density functional theory, our aim is to establish the best trade-off between accuracy and computational efficiency for simulation of atomisitic properties. Motivated by the goal to perform efficient first-principles simulations at an accelerated pace with wider range of applicability, we have leveraged recent development in improved exchange-correlation functionals within the scope of density functional theory.\\ 

Among the plethora of efficient exchange correlation functional schemes within the Generalized Gradient approximation (GGA), meta-Generalized Gradient approximation (meta-GGA), one particular functional - the strongly constrained and appropriately normed (SCAN) approximation stands out. This is a meta GGA functional that satisfies all the 17 exact constraints of a semi-local approximation. The exchange-correlation energy for this functional is given by
\begin{equation}
    E_{xc}^{SCAN}[\rho]=\int dr\rho(r)\epsilon_{xc}^{SCAN}(\rho,\nabla\rho,\tau)
\end{equation}
where \(\rho(r)\) is the electron density, \(\nabla\rho(r)\) its gradient, and \(\tau(r)\) the kinetic energy density \cite{sun2015_SCAN}.  
\\

Density Functional Perturbation Theory (DFPT) is another efficient tool reliant on first principles for the efficient evaluation of the vibrational properties in materials. The perturbative technique is primarily based on the linear response of the properties of the electronic system subject to small displacements in the ionic positions within the lattice. As opposed to the finite difference approach, which requires multiple evaluations of the total energy of the electronic system for different displaced atomic configurations, density functional perturbative approach computes the response of the electronic density and potential to that of small ionic displacements.\\

We utilize density functional perturbation theory to compute the energy obtained by solving self-consistent equations for the first-order variation of the charge density  and the Kohn-Sham potential. For small ionic displacements, the change in the potential is treated as a perturbation to the electronic Hamiltonian.
\begin{equation}
    H(\lambda)=H^{(0)}+V_{ext}(\lambda)
\end{equation}
Here \(\lambda\) is the parameter which defines the small perturbation to the unperturbed Hamiltonian \(H^{(0)}\) due to the effect of the external potential \(V_{ext}(\lambda)\).
The ground state energy of the system is obtained by minimizing the electronic energy functional. 

\begin{align}
    E_{el}[\rho(\lambda)]=\sum_{i=1}^{N_e}\bra{\psi_i(\lambda)}(T+V_{ext}(\lambda))\ket{\psi_i(\lambda)}+E_{H}[\rho(\lambda) \notag\\
    +E_{xc}[\rho(\lambda)]
\end{align}
While the first order derivatives of the energy provides the required forces and dipole moments, second order energy derivatives supplies the dynamical matrix for phonon dispersion relations, and the third derivatives, that of phonon-phonon interactions, Raman modes and Grüneisen parameters.

\subsection{Anharmonic Phonons and Raman-active modes}

The potential energy of a crystal lattice with ionic displacements can be expanded in terms of a Taylor series expansion about the atomic positions at equilibrium. The potential energy corresponding to an atom $\kappa$ in a unit cell $l$ along the direction $\alpha$ is given in terms of the atomic displacement $u_{\kappa\alpha}$ as follows:
\begin{align}
U=U_0+\sum_{\kappa\alpha}\Phi_{\kappa\alpha}u_{\kappa\alpha} + \frac{1}{2!}\sum_{\kappa\alpha, l^{'}\alpha^{'}\beta}\Phi_{\kappa\alpha,l^{'}\alpha^{'}\beta}u_{l\kappa\alpha}u_{l^{'}\kappa^{'}\beta}\notag\\
    + \, \frac{1}{3!}\sum_{l\kappa\alpha, \, l^{'}\kappa^{'}\beta, \, l^{"}\kappa^{"}\gamma}\Phi_{l\kappa\alpha,l^{'}\kappa^{'}\beta, \, l^{"}\kappa^{"}\gamma}u_{l\kappa\alpha}u_{l^{'}\kappa^{'}\beta}u_{l^{''}\kappa^{''}\gamma}+...
\end{align}

The cubic term in this expression incorporates the effect of phonon anharmonicity in the calculations. In terms of bosonic creation and annihilation operators, the third order expansion can be expressed as:
\begin{align}    U_3=\sum_{q\nu,q^{'}\nu^{'},q^{''}\nu^{''}}\Phi_{q\nu, \, q^{'}\nu^{'}, \, q^{''}\nu^{''}}(a_{q\nu}+a^\dagger_{-q\nu})\notag\\
    (a_{q^{'}\nu^{'}}+a^\dagger_{-q^{'}\nu^{'}})(a_{q^{''}\nu^{''}}+a^\dagger_{-q^{''}\nu^{''}})
\end{align}

where $\Phi_{q\nu,\, q^{'}\nu^{'}, \, q^{''}\nu^{''}}$ is the third order force constant, and q and $\nu$ are the phonon wave vector and mode index respectively. In our calculations, multiple supercells were used to generate the displaced atomic coordinates for the evaluation of the third order force constant, whereas all the lower order terms in the above equation can be obtained within the harmonic approximation \cite{phonopy_phono3py}. Our calculation warrants the generation of large number of supercells for improved accuracy of the evaluation of the third order anharmonic terms. Note that, the accurate estimation of the third order force constants are instrumental to the determination of phonon lifetime and phonon spectral functions.\\

The phonon lifetimes can be expressed as:
\begin{equation}
    \tau_{q\nu}=\frac{1}{2\Gamma_{q\nu}(\omega_{q\nu})} 
\end{equation}
where $\Gamma_{q\nu}$ is the imaginary part of the phonon self energy. The effect of interactions is included in the self-energy resulting in finite lifetimes of the phonon modes. More often, the phonon lifetimes are dependent on changes in the temperature of the lattice. In an effort to account for the above, a temperature dependent effective potential (TDEP) has been implemented to incorporate the higher order (fourth) displacements in the Hamiltonian to successfully elucidate the anharmonic phonons in certain halide double perovskite materials \cite{Klarbring2020}. In systems with non-negligible anharmonicity, the frequency of phonon modes renders a significant change with temperature, and it's signatures are evident from the thermodynamic properties of the material. In such cases, the quasiharmonic approximation (QHA), which accounts only for the change in volume caused by thermal expansion, might not be adequate. This is primarily due to the fact that the intrinsic anharmonic effects also play an important role at higher temperatures \cite{Zhang2019}. \\

At the microscopic level, the vibration of the atoms affect the distribution of electronic charge which modulates the polarizability of the system. The components of the electronic polarizability tensor are denoted by $\alpha_{ij}$. For a crystalline system where the polarizability is intrinsically dependent on the normal mode coordinate $Q_{q\nu}$, the motion of the normal mode Q couples with an incident photon and that of the scattered radiation, resulting in a shift in the frequency. Such a shift in the frequency can be detected by Raman spectroscopy and the components of the Raman tensor, $R_{\nu}$ are defined as: 
\begin{eqnarray}
    R_{\nu,ij}=\left.\frac{\partial\alpha_{ij}}{\partial Q_{\nu}}\right|_{Q=0}
\end{eqnarray}

The Raman scattering intensity of a phonon mode is proportional to $|R_{\nu}|^2$. The experimentally observed intensity of the Raman spectrum strongly relies on the polarization of the incident light and the symmetry of the Raman tensor. Note that, for a particular phonon mode to be Raman active, the value of Raman tensor has to be non-zero. This condition on the phonon mode symmetry is decided based on group theory. \\
 \begin{figure}[H]
     \centering
     \includegraphics[width=1.0\linewidth]{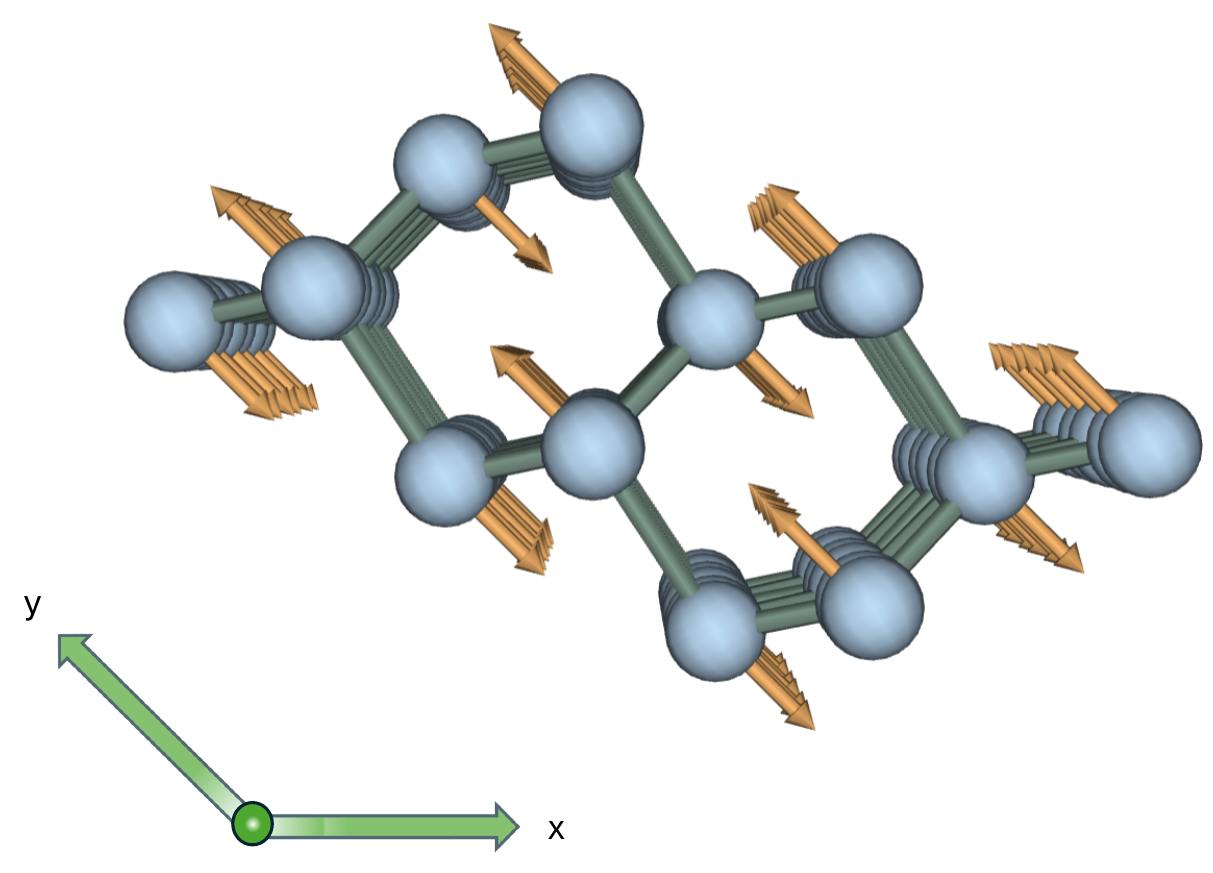}
     \caption{Schematic diagram of Raman active phonon vibration for 2-H silicon. }
    \label{fig:placeholder}
 \end{figure}
 
The phonon modes at the center of the Brillouin zone of a crystal transform following certain irreducible representations of the point group of the crystal. A phonon mode is said to be Raman active if the symmetry of the mode matches one or more components of the Raman tensor allowed by the point group. The direct product of the irreducible representation of the mode with the polarizability tensor must be included in the symmetry. If we consider acoustic phonon modes in the zone center, the polarizability of these modes do not change because of simple translation of the lattice. This is why acoustic modes are Raman inactive. On the other hand, the optical modes involve motion of atoms in opposite directions within the unit cell and can cause a change in the polarizability. These modes contribute to Raman spectra if they are allowed by symmetry considerations. This reflects the fact that the spectra is highly sensitive to symmetry of the system.\\

The connection between Raman activity and phonon eigenvectors becomes evident particularly in the light of first-principles treatment. In density functional perturbation theory (DFPT), Raman tensors are directly computed as third derivatives of the total energy with respect to one atomic displacement and two sets of electric fields, or in other words, derivatives of the macroscopic dielectric tensor with respect to phonon coordinates \cite{baroni2001phonons}. This approach incorporates both electronic and vibrational contributions and allows the prediction of Raman frequencies and intensities without the need for empirical parameters. The method, though successful in it's own regime, comes with the demerit of the inability to evaluate the resonant Raman spectra. Nevertheless, the above first-principles Raman calculations have become the prototype for interpreting experimental spectra and for assigning observed Raman peaks to specific phonon modes in functional materials. Figure 1 displays the structure of the prototypical hexagonal silicon and germanium polymorphs, the arrows indicate the vibrational phonon modes for the Raman active optical modes close to the center of the Brillouin zone as visualized from the z direction of the crystal lattice.\\

Ab-inito formalisms based on Density functional perturbation theory have been recently developed for the computation of first order resonance Raman spectra of structures of quantum materials obtained from first principles \cite{Hung2023}. We can obtain either the Raman excitation spectra, which maps the variation of the Raman intensity for a particular phonon mode with the frequency of incident laser, or the Raman spectra which measures the intensity versus shift in energy of the incident photons. The inverse of the linewidth of the former peaks provide us with the lifetime of the photo excited electrons, and the latter supplies the phonon lifetimes. If circularly polarized light (CPL) is the source radiation, it is feasible to additionally investigate the helicity dependent Raman spectra as well. Certain phonon modes conserve the helicity after scattering and thereby indicates the component of the total angular momentum of a photon in the direction of propagation while other modes do not. Such a quantitative analysis provides deeper insight into the symmetry of the phonon modes.\\

For circularly polarized light, the polarization vector such that $P_x=P_y$ and $\phi=\frac{\pi}{2}$ corresponds to left handed polarization ($\sigma_+$), while that of $\phi=\frac{-\pi}{2}$ corresponds to right handed polarization ($\sigma_-$). The two polarization vectors are definend as follows:
\begin{eqnarray}
    P_{\sigma_+}=\frac{1}{\sqrt{2}}\begin{pmatrix}
        1\\
        i\\
        0
    \end{pmatrix}\\
    P_{\sigma_-}=\frac{1}{\sqrt{2}}\begin{pmatrix}
        1\\
        -i\\
        0
    \end{pmatrix}
\end{eqnarray}
Incorporation of these polarization vectors into the expression for Raman tensor leads to the helicity dependent spectra for a circularly polarized light. We have computed the expression for Raman intensity from third order perturbation theory utilizing ab initio techniques by evaluation of the electron-photon and electron-phonon matrix elements. 

\subsection{Transport and thermoelectric properties}

The canonical partition function for a system of non-interacting phonons under the harmonic approximation is given by:
\begin{equation}
    Z=\prod_{q\nu} [ \frac{e^{-\hbar \omega _{q\nu}/2k_BT}}{1-e^{-\hbar\omega_{q\nu}/k_BT}}] 
\end{equation}
where $\omega_{q\nu}$ is the phonon frequency corresponding to mode $\nu$ and wave vector $q$, $k_{B}$ the Boltzmann constant and T the temperature of the phonons. We can obtain the Helmholtz free energy from the partition function using the standard expression $F=-k_{B}T$ln$Z$, resulting in the following equation:
\begin{equation}
    F=\frac{1}{2}\sum_{q\nu}\hbar \omega_{q\nu}+k_BT\sum_{q\nu}ln[1-e^{-\hbar\omega_{q\nu}/k_BT}]
\end{equation}

This leads to the following expression for entropy obtained from the partition function:
\begin{align}
    S=\frac{1}{2T}\sum_{q\nu}\hbar\omega_{q\nu}coth(\hbar\omega_{q\nu}/2k_BT)\notag\\
    -k_B\sum_{q\nu}ln[2sinh(\hbar\omega_{q\nu}/2k_BT]    
\end{align}

Further, the heat capacity at constant volume is evaluated from the energy of the harmonic phonons, and is given as:
\begin{equation}
    C_v=\sum_{q\nu}k_B(\frac{\hbar\omega_{q\nu}}{k_BT})^2\frac{e^{\hbar\omega_{q\nu}/k_BT}}{[e^{\hbar\omega_{q\nu}/k_BT}-1]}
\end{equation}

The above quantities are evaluated by sampling of the phonon modes over a uniform $q$-grid in the first principles simulations followed by performing the required summation over all modes \cite{phonopy_phono3py}.
\subsection{Temperature dependent mode Grüneisen Parameters}

The frequencies of the phonon modes are most often observed to vary with a small change in volume of the unit cell. The variation usually exhibits a nearly linear behavior. This change is characterized by a dimensionless quantity named the mode Grüneisen parameter ($\gamma$).
The parameter determines the contribution of the phonon modes to the thermal expansion, quantifies the extent of phonon-phonon scattering and the softening of phonon modes as a function of temperature \cite{fabian1997thermal}, \cite{lee2017anharmonic}. The Grüneisen parameter is of utmost importance for quantifying lattice anharmonicity and serves as a crucial descriptor for machine-learning based phonon spectrum analysis \cite{cuffari2020calculation}, \cite{toher2014high}.  
The most widely prevalent expression for Grüneisen parameter is as follows:
\begin{equation}
    \gamma _i (q)=\frac{\partial[ln \omega_i (q)]}{\partial (ln V)}
\end{equation}
here $\omega_i(q)$ is the frequency of $i^{th}$ phonon mode, with wave vector being $q$ and V, the lattice unit cell volume. \\
An alternative expression often calculated for pratical purposes is the weighed average of the Grüneisen parameter:
\begin{equation}
    \gamma_{avg}=\frac{\sum_ql\gamma_i(q)C_i(q)}{ \sum_{q} C_i(q)}
\end{equation}
Here $C_i$ is the specific heat of $i^{th}$ phonon mode. Note that, in reference to the above expression, a larger magnitude of the averaged Grüneisen parameter, is indicative of higher lattice anharmonicity. Similarly, a larger contribution to this parameter from certain modes, would suggest that they are mostly responsible for their contribution to thermal conductivity in the material.
It is often possible to incorporate explicit temperature dependence in the evaluation of thermodynamic Grüneisen parameter by implementing the temperature dependent effective potential (TDEP) method \cite{Hellman2013}. In an effort to account for the above, the third order interatomic force constant have been computed via the least squares technique in our formalism to calculate the detailed structure of the Grüneisen parameters \cite{Gonze2020}. 
\section{Results and Discussion}
\subsection{Electronic structure and charge density calculations}

First principles electronic structure and phonon calculations were performed using the projector augmented-wave (PAW) method under the Perdew-Burke-Ernzerhof (PBE) exchange-correlation functional scheme as implemented in the Quantum espresso package. The structural lattice parameters of hexagonal 2-H lonsdaleite silicon considered for our first principles calculations are $a=b=$ 3.83 \AA, and $c=$ 6.63 \AA. This is in close agreement with the experimentally synthesized structural dimensions of $a=$ 3.8242 \AA  \: and $c =$ 6.3237 \AA \cite{hex_si_realized}. While the generalized gradient approximation (GGA) implementation improves prediction of electronic structure for a multitude of materials due to the incorporation of the density gradients in the functional, it is known to underestimate band gaps in certain cases because of self-interaction errors and the absence of derivative discontinuity. As a result, advanced approaches to estimation of functionals such as meta-GGA or hybrid functionals have often been utilized to remedy these shortcomings and have been successful in accurate predictions of materials properties. However, these accurate formalisms such as the hybrid functionals are computationally quite expensive, and hence are not the best choice of tradeoff between computational efficiency and accuracy.\\ 

Starting from the lattice parameters stated above, we have performed detailed structural optimization for finding the most stable configuration. Our ab initio results yield the relaxed lattice parameters for hexagonal silicon to be given by $a=$ 3.850 \AA, $c=$ 6.365 \AA (c/a = 1.653). Note that, the in-plane lattice constant has increased by only $\sim$ $0.5\%$ whereas, the out of plane lattice parameter has reduced by over $\sim$ $4\%$. The starting lattice parameters for lonsdaleite germanium in our calculations are $a=b=$ 3.997 \AA, and $c=$ 6.599 \AA. Our first principles results yield the relaxed lattice parameters for hexagonal germanium to be given by $a=$ 3.978 \AA, $c=$ 6.696 \AA (c/a = 1.658). Note that, the in-plane lattice constant has decreased by $\sim$ $0.48\%$ whereas, the out of plane lattice parameter has only decreased by over $\sim$ $0.04\%$. \\

A convergence of the electronic bands was obtained with respect to both the $k$-point grid as well as the energy cutoff for the plane wave basis sets. Convergence was achieved at the energy cutoff of 45 Ry and using a $10\times10\times5$ $k$-point grid, with a denser $12\times12\times6$ $k$-point grid for the non-self-consistent calculations. The wavefunctions are expanded in a basis set of plane waves and we have utilized the pseudopotential method such that the $3s$ and $3p$ electrons are in the valence shell for silicon, whereas the $3d$, $4s$ and $4p$ electrons are in the valence for germanium.\\

One of the most fundamental requisites for device applications of quantum materials is their band gap alignment and tuning of the band gap by strain and alloy formation. Motivated by the controversies in band gap estimations, we have evaluated the band gaps of silicon and germanium utilizing the popular Generalized Gradient Approximation (GGA) functional. A detailed comparison of the band gaps of the hexagonal phases of silicon and germanium till date have been presented in Table I. Note that, the previously reported value of band gap obtained using GGA for lonsdaleite silicon by Fan et. al. \cite{Chen2017} is $\simeq$ 0.45 eV, whereas those obtained from the advanced GW approximation is $\simeq$ 0.95 eV. Utilization of the more accurate Hybrid functional HSE (Heyd, Scuseria, and Ernzerhof) calculations predict a similar indirect band gap of $\simeq$ 0.984 eV, with the conduction band minimum at M, consistent with previous literature \cite{keller2023first}. Fig. 2 displays the electronic structure and density of states for both silicon and germanium compounds. Our calculations for hexagonal silicon utilizing the relaxed lattice parameters reveal an indirect band gap of 0.419 eV, similar to the previous calculations with a band gap difference of within $\simeq$ 7.5\% due to the effect of choice of lattice parameters, basis sets, and pseudopotential methods. Simultaneously, we have utilized the famous Strongly Constrained and Appropriately Normed (SCAN) functional, recently observed to predict improved band gaps and functional characteristics for a wide range of material phases. First principles calculations employing the SCAN functional estimates a band gap of 0.769 eV, which agrees with band gaps predicted utilizing the $k.p$ theory. \\


Next, we explore the nature of the band gap and low energy properties of the sister compound, 2-H lonsdaleite germanium. The relaxed lattice parameters from our converged first-principles calculations are within $\simeq$ 2.5\% of the experimental lattice parameters. While, the popular GGA functional erroneously predicts the hexagonal germanium to be a metal, advanced functionals aka the Hybrid functional and the Green's formalism yields band gaps to 0.283 and 0.23 eV respectively. Interestingly, lonsdaleite germanium, as opposed to it's silicon counterpart, has been reported to host a direct band gap, making this material, and it's alloys, potential candidate for photovotaic and optoelectronic applications. First principles calculations utilizing the semi-empirical meta-GGA exchange-correlation functional of the modified Beche-Johnson potential (mBJ) and $k.p$ theory estimates slightly larger band gap of $\simeq$ 0.31 eV. The detailed comparison of the band strutures and nature of the gap for both the hexagonal silicon and germanium compounds have been presented in Table I. Note that, even though, the experimental direct band gap in germanium has been reported to be 0.35 eV, the measurements are done using photoluminescence spectroscopy, thus, some of the optical transition matrices are forbidden by the selection rules disqualifying some of the bands near the band edges and the actual band gap could, in all probability, be smaller than the reported value. The actual band gap can be verified by transport measurements. We not that, albeit the fact that, hybrid functional calculations are accurate for a multitude of atomistic simulations, it is plagued by inaccuracies for metals and very small band gap systems, and high computational costs leading to decreased computational efficiency. The GW methodology, on the other hand, predicts band gaps larger than experimental values due to the neglect of vertex corrections and underestimated screening effects.\\

In order to perform first principles calculations of the band structure for hexagonal germanium, we have employed the Perdew Burke Ernzerhof (PBE) formalism of the GGA exchange-correlation functional. The calculations were converged with respect to the energy cut-off of 40 Ry, and for a $k$-point sampling grid of $8\times8\times4$. Our estimation of band structure utilizing ab-initio simulations do not reflect any visible band gap, either direct or indirect, in conformity with previously predicted results. Apart from the hybrid functionals and GW approximation, the relatively less cumbersome meta-GGA SCAN functional shows immense potential in accurate prediction of electronic structure in a variety of materials with different atomic environments and phases\cite{Pokharel2022}. Here, we have performed ab-initio calculations utilizing the SCAN functional, which is known to outperform GGA and even hybrid functional calculations in recent years, even though it requires, at best, only a fraction of the computational cost. From pur investigation, utilization of the SCAN meta GGA functional estimates a direct band gap of 0.2 eV for 2-H germanium.  Fig. 2 displays a comparison of the band structure and density of states for silicon and germanium corresponding to the GGA and meta-GGA SCAN functionals. Note that, there is a pronounced difference between the valence bands obtained from the GGA and the meta-GGA SCAN functionals, particularly for the case of silicon. Further, there is a significant hybridization of the bands near the band edges in the dispersion relations at the commensurate $\Gamma$ point  in the case of the germanium polymorph. A closer look at the density of states reveals a significant concentration of available states accessible for transitions near the band edges in the solar spectrum. The possibility of multiple electron-hole pair generation combined with the direct band gap deems the 2-H germanium polytype, an ideal candidate for optoelectronic applications.\\

In an effort to compare prevalent exchange-correlation functionals in terms of the accuracy and computational efficiency, we note that, the hybrid functional calculations have been deemed inaccurate in certain systems due to the incorporation of exchange interactions within the Hartree-fock theory, and the GW approximation neglects vertex corrections to the self energy, while treating screened Coulomb interactions. More importantly, these methods could result in more accurate results compared to the GGA exchange correlation functional, however, their weakness lies in the computationally efficient of prototypical quantum simulations, especially of large scale atomistic systems. Albeit some discrepancies in the accurate estimation of magnetic moments utilizing the SCAN functional, the meta-GGA formalism provides by far the best tradeoff between accuracy and computational efficiency till date. Figure 3 displays a comparison of the charge density plots within the unit cell of hexagonal silicon performed utilizing the SCAN meta-GGA functional with a converged 12 $\times$ 12 $\times$ 6 k-point mesh. The inset of Figure 3 shows the difference in the electronic charge densities estimated with that of SCAN and GGA functionals,    $\delta \rho_{xc}$ = $\rho^{SCAN}_{xc}-\rho^{GGA}_{xc}$. Our charge density calculations succeed in encapsulating the formation of bonds and electron localization accurately for hexagonal silicon. We note that, being a moderately correlated system with relatively reduced self interaction contribution, hexagonal silicon and germanium are among the ideal testbed candidates for the improved SCAN meta-GGA functionals. Interestingly, the incorporation of the kinetic energy density $ \tau(\vec{r})=\frac{1}{2}\sum_{i=1}^{occ}\left | \nabla \psi_{i} \right |^{2}$ and the use of semi-local exchange correlation functional is apparent from the inset plot of the charge density difference of the two functionals. Note that, the meta-GGA functional provides a refined charge density contour mapping that replicates the localization and electron distribution along the directions of the $Si-Si$ bonds.\\ 
\begin{figure}[h]
    \centering
    \includegraphics[width=1\linewidth]{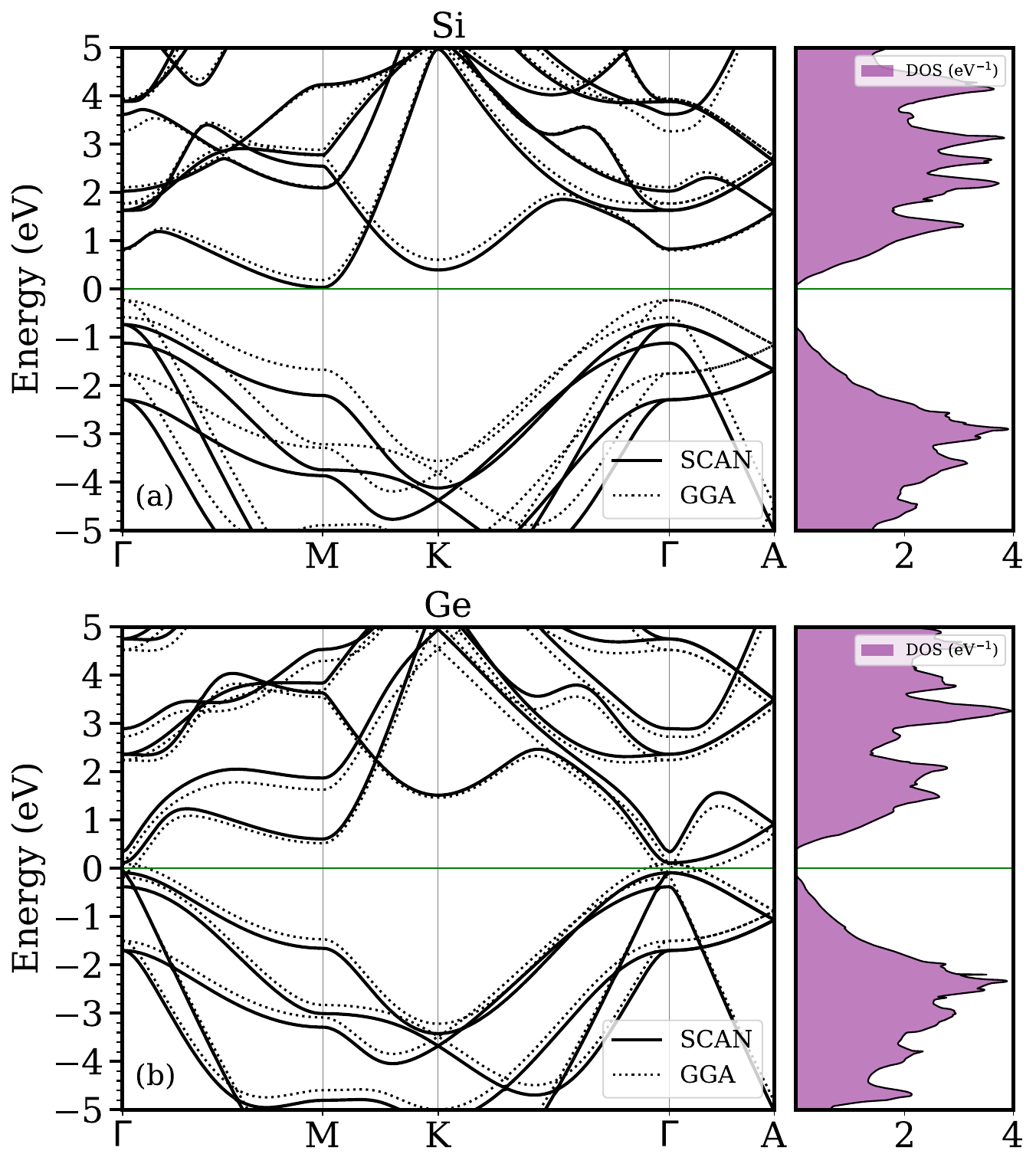}
    \caption{Electronic band structures and corresponding density of states of hexagonal silicon (a) and hexagonal germanium (b). Solid lines represent SCAN meta-GGA results, while dotted lines correspond to GGA calculations. }
    \label{fig:enter-label}
\end{figure}
\begin{figure}[htbp]
    \centering
    \includegraphics[width=1\linewidth]{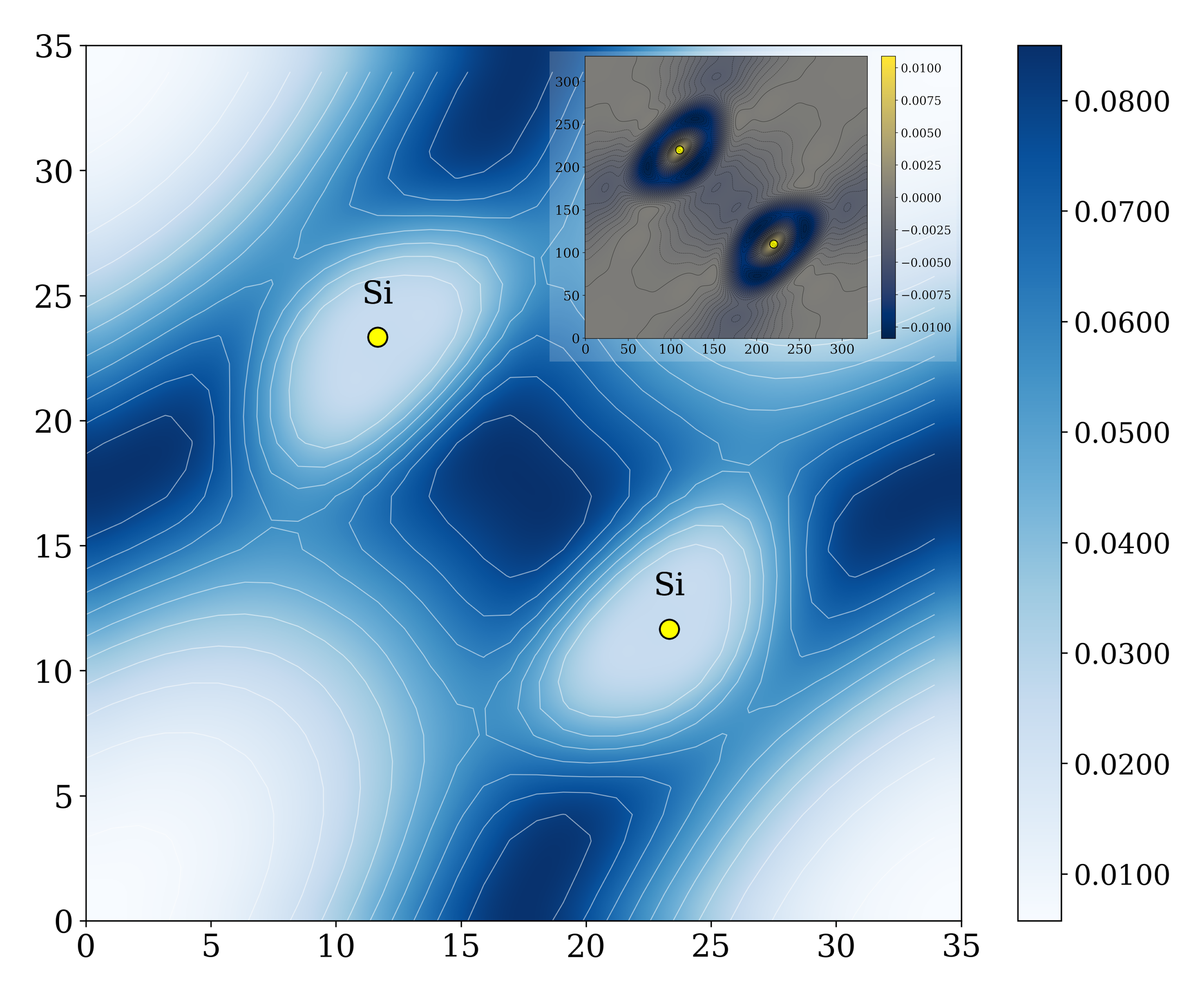}        
    \caption{The electronic charge density for 2-H silicon under GGA formalism, with the charge density difference between SCAN and GGA in the inset.}
    \label{fig:placeholder}
\end{figure}
\begin{table}[htbp]
\caption{\label{tab:table2}
Bandgap hexagonal silicon and germanium computed with different exchange correlation functionals.}
\resizebox{\columnwidth}{!}{\begin{tabular}{|c|c|c|l|l|}
\hline
Polymorph& Methodology&a (\AA)&c (\AA)& Band gap\\
\hline
Si&GGA (Our work)&3.85&6.365&0.419 ($\Gamma\xrightarrow{}$M)\\

&SCAN (Our work)&3.85&6.365&0.769\\

 &GGA (Fan et al.) &3.848&6.380&0.45\footnotemark[4]\\
 &GW&3.828&6.325&0.95\footnotemark[6]\\

 &Empirical model &3.836&6.264&0.796\footnotemark[3]\\
  &HSE06&3.826&6.327&0.984\footnotemark[7]\\
 \hline
Ge&GGA (Our work)&4.063&6.686&0.0\\
&SCAN (Our work)&4.063&6.686&0.202 (Direct)\\
 &GGA (Fan et al.)&4.036&6.671&-\footnotemark[4]\\
 &Empirical model&3.993&6.520&0.310\footnotemark[3]\\

 &GW&3.96&6.45&0.23\footnotemark[5]\\
&HSE06&3.993&6.589&0.283\footnotemark[7]\\
 &Experiment (Optics)&3.960&6.570&0.350\footnotemark[2]\\\hline
\end{tabular}}
\footnotetext[1]{Reference \cite{Cai2021}.}
\footnotetext[2]{Reference \cite{hexge_expt_bg}.}
\footnotetext[3]{Empirical model pseudopotential, reference \cite{De2014}.}
\footnotetext[4]{Reference \cite{Fan2018}.}
\footnotetext[5]{Reference \cite{Chen2017}.}
\footnotetext[6]{Reference \cite{Rdl2015}.}
\footnotetext[7]{Reference \cite{keller2023first}.}
\end{table}
\subsection{Phonon dispersion relations and Raman mode decomposition}

 An accurate estimation of the phonon dispersion relations is pivotal to determining the stability and their electronic signatures such as soft modes, raman spectroscopy, anharmonicity and phonon-phonon interactions. In this section, we elucidate the phonon dispersion relations obtained for both hexagonal silicon and germanium evaluated using both  density functional perturbation theory as well as the supercell approach. The results from these two computational methodologies are in agreement with each other. Figure 4 displays the phonon dispersion relations obtained from our first principles calculations for the silicon and germanium allotropes. The decomposition of the Raman modes for the different phonon modes are displayed in separated color codes. Note that, apart from the acoustic modes, the phonon plots are highly dispersive in the 6 - 13 THz range of the spectral decomposition. The optical phonon modes in the 6 - 13 THz range for silicon and 3 -8 THz for gemanium correspond to lower density of states compared to the acoustic modes. However, there is a significant increase in the density of states of the optical phonon mode around 13 - 15 THz (62.1 meV) for  2-H silicon and 8 - 10 THz (37.26 meV) for 2-H germanium. The decomposition of the phonon spectra reveals that the highly localized and nearly flat phonon bands in the 13 - 15 THz for silicon, and  8 - 10 THz for germanium of the spectra constitute the Raman active modes, $A_{1g}$, $E_{1g}$ and $E_{2g}$ respectively. Despite having similar features, the difference in frequency range of the phonon spectra of both materials is associated with the relatively heavier mass of Ge atoms which reduces the vibrational frequencies \cite{borlido2026ab}. We note that imaginary phonon frequencies are not observed for either of the materials, indicating the fact that the optimized structures are predicted to be stable from first principles calculations. 
\begin{figure}[H]
    \centering
    \includegraphics[width=1\linewidth]{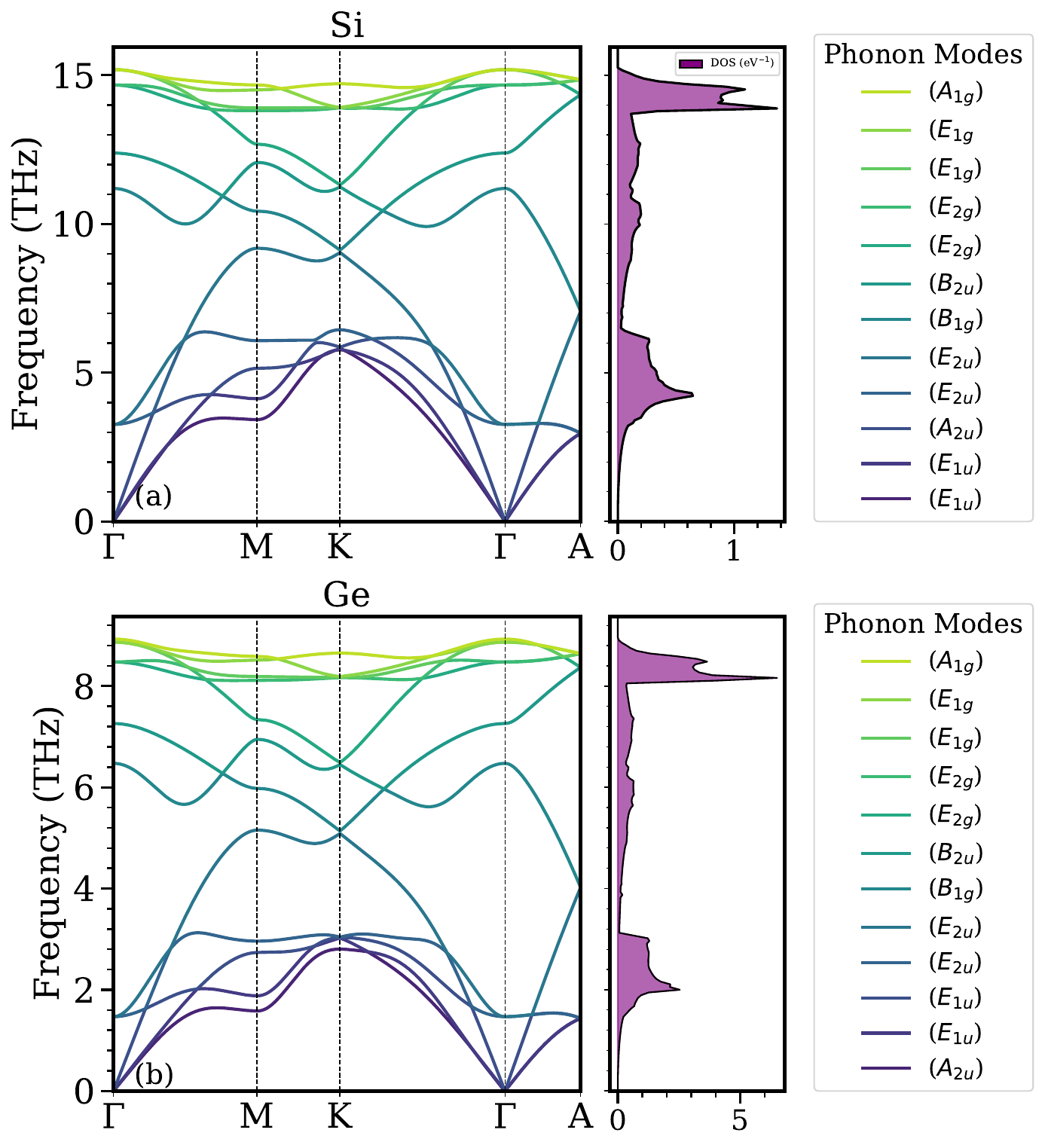}
    \caption{Phonon dispersion and density of states for hexagonal Si (a) and hexagonal Ge (b) with the symmetry of the phonon modes}
    \label{fig:enter-label}
\end{figure}
It has been observed that relaxation of the structure leads to symmetry breaking in hexagonal germanium, resulting in phonon modes of different symmetry than in the non-relaxed one which gives phonon modes that agree well with conventional hexagonal systems. Group theory provides the mathematical framework to classify the normal vibrational modes into irreducible representations of the crystal’s point group, which describe how each mode transforms under the symmetry operations of the crystal structure. Raman spectroscopy is a powerful technique for probing the symmetry properties of vibrational modes in a crystal. A vibrational mode is Raman-active only if it induces a change in the polarizability of the unit cell; the mode must have the same symmetry as any component of the polarizability tensor. In the case of hexagonal structures, group-theoretical analysis predicts three Raman-active phonon modes at the center of the Brillouin zone as the $A_{1g}$ mode, which is a non-degenerate longitudinal optical vibration, and the $E_{1g}$ and the $E_{2g}$ modes, which are doubly degenerate transverse optical vibrations. These Raman-active modes are directly observable in the Raman spectrum and serve as signatures of the crystal’s vibrational symmetry\cite{hex_si_realized}.\\
\

We note that for the lonsdaleite compounds crystallizing in the $P63/mmc$ space group with $D_{6h}^4$ symmetry, the Raman modes $A_{1g}$, $E_{1g}$ and $E_{2g}$ should indeed be Raman active from symmetry analysis.\cite{Aroyo:xo5013} Here, the notations A and E modes refer to Raman modes which are symmetric with respect to the principle axis of symmetry, and doubly degenerate in the two dimensional irreducible representations respectively, while the subscripts 1 and 2 refer to whether they are symmetric or antisymmetric with respect to the $C_{2}$ axis perpendicular to the principle axis of symmetry for this material.\\

Figure 5 displays the relative Raman intensity peaks as a function of Raman shift for 2-H silicon and germanium. Our first principles calculation for hexagonal silicon, result in strong Raman-active modes observed at 496 $cm^{-1}$, and 468 $cm^{-1}$. As for 2-H Ge, the corresponding Raman intensity peaks are observed at 276.18 $cm^{-1}$ and 261 $cm^{-1}$ respectively. The modes with higher relative intensity characterized by a low depolarization ratio of 0.35, corresponds to two Raman active modes, $A_{1g}$ out of plane vibration and doubly degenerate $E_{1g}$, whereas the doubly degenerate modes with depolarization ratio 0.75 arises from in plane E type vibrations of $E_{1g}$ mode for both systems. The peaks from the $A_{1g}$ and $E_{1g}$ modes, despite being Raman active, are too closely spaced and hence  were not resolved separately in the spectrum, a feature which has been confirmed by previous first -principles calculations as well. \cite{borlido2026ab}\\

\begin{table}[H]
    \centering
    \resizebox{\columnwidth}{!}{\begin{tabular}{|c|c|c|}
    \hline
    Polymorph&Raman shift (cm$^{-1}$)&Avg. Grüneisen parameter $\gamma$\\
    \hline
   2-H Si&496&1.006 ($A_{1g}$, $E_{1g}$)\\
   \hline
   &468&0.944 ($E_{2g}$)\\
   \hline
   2-H Ge&276&1.04($A_{1g}$, $E_{1g}$) \\
   \hline
   &261&1.368 ($E_{2g}$)\\
   \hline
   
   \end{tabular}}
    \caption{Raman-active phonon frequencies and their corresponding vibrational modes obtained from first-principles calculations. The third column reports the average Grüneisen parameter associated with each Raman-active phonon mode.} 
    \label{tab:placeholder}
 \end{table}
   
The helicity-dependence of the resonant Raman spectra for circularly polarized light  can be evaluated depending on whether the Raman modes conserve or change the helicity of the right or left circularly polarized light by evaluating the symmetry of the Raman tensor and change in angular momentum of the photon. The first order Raman intensities have been evaluated by incorporating the electron-photon and electron-phonon coupling matrix elements computed from ab initio calculations \cite{Hung2023}. Certain phonon modes conserve the helicity after scattering while others do not. Note that, the helicity refers to whether the incident and scattered photon sources are right or left handed, as it indicates the component of the total angular momentum of a photon in the direction of propagation.\\

The expression for the Raman intensity as per the third order perturbation theory, which involves the system interacting through three perturbative terms before reaching the final state, is given by: 
\begin{align}
\begin{split}
I&(E_{L},E_{R})  = \\
&\sum_{\nu} \left|\sum_{k,i,n,n^{'}}\frac{M^{opt}_{n^{'}\to i}(k)g^{ep}_{n\to n^{'}}(k, \nu)M_{i\to n}^{opt}(k)}{[E_{L}-\Delta E_{ni}(k)][E_{L}-\Delta E_{n^{'}i}(k)-\hbar\omega_{\nu}]}\right|^2 \\
& \times \delta(E_{R}-\hbar\omega_\nu)  
\end{split}
\end{align}
The above equation provides an expression for the scattering probability as a function of the energy of the incident photon source $E_L$, and the corresponding Raman shift $E_R$.  Here, $i$ and $f$ denotes the initial and final electronic state, and $n$ and $n^{'}$ represent the virtual electronic states involved in the scattering process. Note that the final state $f = i$ for the electron-hole recombination process. The electron-photon matrix elements $M_{i\to n}^{opt}(k)$, and $M_{n^{'}\to i}^{opt}(k)$ incorporate the coupling between the electronic states during the photon absorption and emission processes respectively. The electron-phonon matrix elements $g^{ep}_{n\to n^{'}(k,q)}$ quantify the scattering between the intermediate states $n$ and $n^{'}$ mediated by electron-phonon interaction involving the phonon mode with momentum $q$ and frequency $\nu$. The first order Raman scattering process results from the inelastic scattering of the electronic states involving a single vibrational mode. As per the momentum conservation requirements, only zone center phonon modes at the $\Gamma$ point $(q\approx 0)$ take part in first order scattering. The terms $\Delta E_{ni}$ and $\Delta E_{n^{'}i}$ in the denominator of the equation denote the difference in energies between the electronic states incorporating a small broadening parameter to account for the finite lifetime of the excited electronic states. The delta function $\delta (E_R-\hbar \omega_\nu)$ ensures that the Raman shift agrees with the phonon energy. This term is replaced by a Lorentian function to include phonon lifetimes in the practical implementation. \\

The electronic band structure plays a pivotal role in understanding the amplitude of the Raman scattering intensities. Note that the intermediate electronic states between the excitation and recombination processes decide the amplitude of the scattering. If the energy of the incident photon source is almost identical with that of the electronic energy level difference, that would result in a spike in the intensity of Raman spectra at the corresponding frequency, thereby resulting in the resonant Raman scattering process \cite{cardona2006light}. The above statement validates the involvement of both the phonon modes and that of the electronic structure in contributing towards the observation of Raman active modes in the material of interest. Therefore, the analysis of the Raman modes indeed sheds light into a deeper understanding of the decomposition of the phonon spectra and their involvement in phonon-assisted optics.

\begin{figure}[h]
    \centering
    \includegraphics[width=1\linewidth]{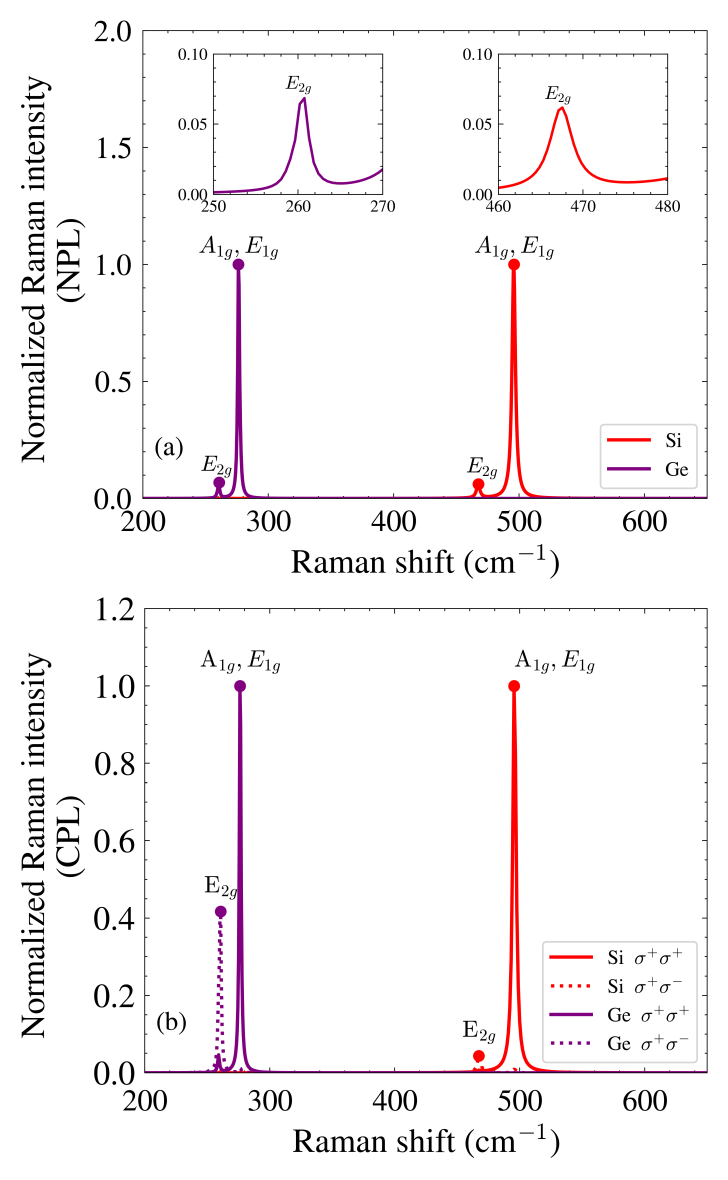}
    \caption{Resonant Raman spectra of hexagonal Si and Ge computed at an excitation energy of 2.33 eV (a), with the spectra for non-polarized light with weak modes enlarged in the insets, and the helicity-resolved Raman response under circularly polarized excitation (b).}
    \label{fig:placeholder}
\end{figure}

Figure 5 displays the resonant Raman spectra as a function of the Raman shift for non-polarized and circularly polarized light. The calculated Raman spectral distributions of hexagonal Si and Ge present two prominent peaks of Raman-active optical phonon modes at the zone center. Figure 5 (a) shows the Raman spectra for a non-polarized incident photon source. The corresponding Raman spectra consists of two asymmetric spectral features for both lonsdaleite
Si and Ge. The amplitude of the smaller peak is approximately 15 times lower than the larger peak observed from calculations of Raman intensity utilizing equation (16). While the smaller peak is predominantly of $E_{2g}$ symmetry, the larger peak is a superposition of two closely placed peaks of $A_{1g}$ and $E_{1g}$ symmetry. We infer from our results displayed in Figure 5 as well as the tabulated values from Table II, that the $E_{2g}$ and $A_{1g}+E_{1G}$ peaks occur at 468 $cm^{-1}$ and 496 $cm^{-1}$ respectively for Si and 261 $cm^{-1}$ and 276 $cm^{-1}$ respectively for Ge. Hence, the difference in the two peaks for 2-H Si is almost twice compared to that for 2-H Ge. Further, the Raman shifts for phonon frequencies of Ge are observed at much lower values compared to Si because of the relatively heavier atomic mass of Ge.\\

Next, we discuss the results for the circularly polarized light source as depicted in Figure 5 (b). The $E_{2g}$ mode which is doubly degenerate is found to change the helicity of the circularly polarized light as per the conservation of angular momentum \cite{Tatsumi2018}. This is observed in the Raman spectra for both the materials, Si and Ge. Interestingly, the Raman peak composed of $A{1g}$ and $E_{1g}$, on the other hand preserves the helicity of light contrary to the other peak. We note that the Raman frequencies obtained utilizing the DFT-LDA (local density approximation) functionals are slightly lower than the values experimentally measured and previously published\cite{Ahn2021}. The above is a known artifact of performing phonon calculations utilizing first principles within the harmonic approximation at absolute zero temperature. Albeit the fact that the position of the Raman peaks remain unchanged under circularly polarized light, there is a significant change in the spectral weights of the Raman intensities. The helicity changing Raman mode has larger spectral weight by more than twice for 2-H Si, whereas the helicity preserving mode dominates the spectral distribution by the same order, in the case of 2-H Ge. \\

While the Raman active modes provide valuable information about the phonon modes and their assistance in scattering of non-polarized and polarized light, it is crucial to study the phonon lifetimes and linewidths using density functional perturbation theory. Figure 6 below depicts the plots of phonon lifetimes as a function of the phonon frequencies for hexagonal silicon and germanium at different temperatures. 

\begin{figure*}[t]
    \centering
    \includegraphics[width=\textwidth]{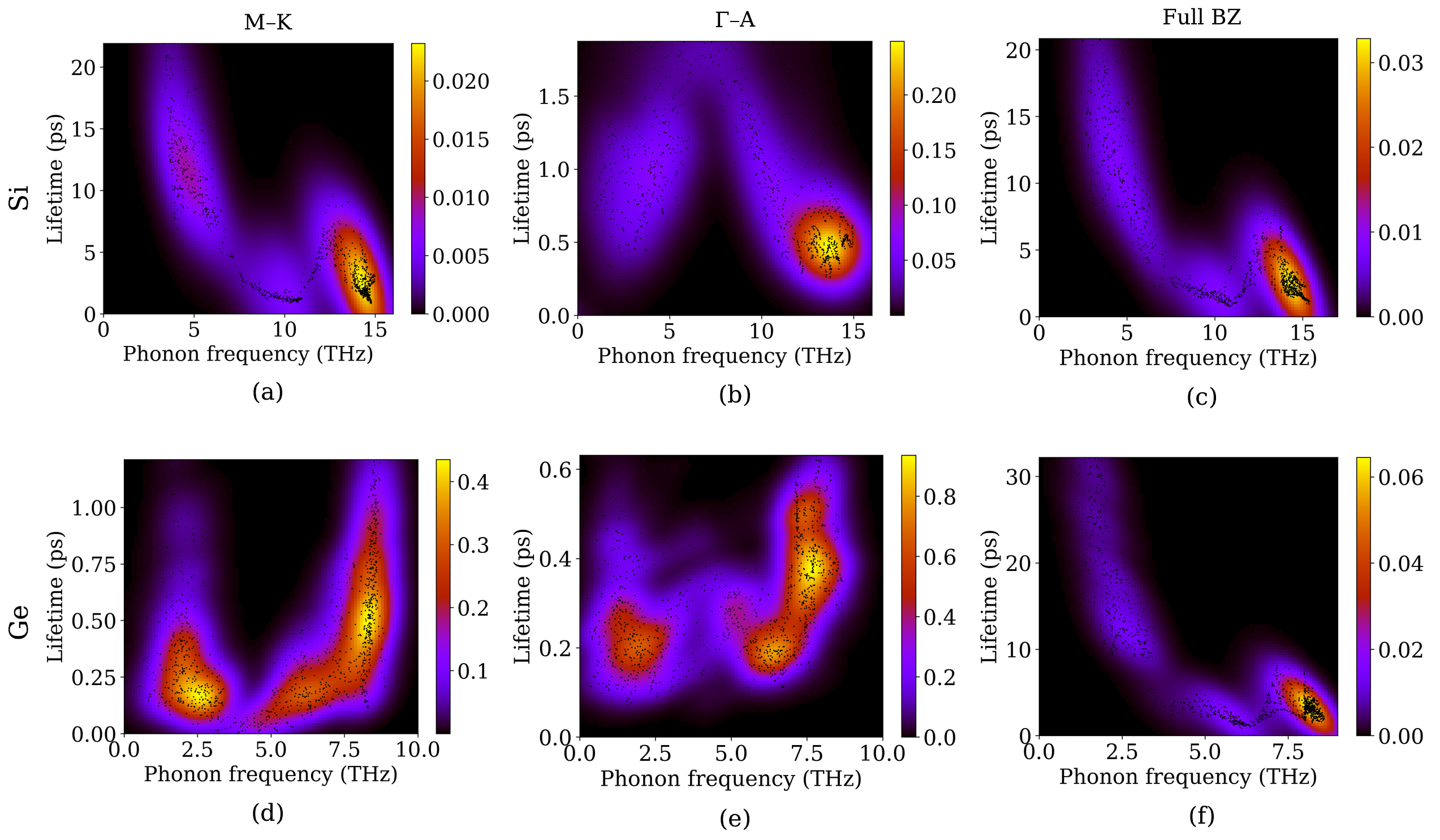}
    \caption{Phonon lifetime of 2-H Si and Ge for the M-K and G-A symmetry paths and the full Brillouin zone at 300 K.}
    \label{fig:enter-label}
\end{figure*}

The plots associated with Fig. 6 are indicative of the ranges of phonon lifetimes, whereas the heatmap represents the density of the distribution of phonon modes. The phonon density distributions are evaluated using Gaussian kernel density estimation (KDE) function. This technique estimates the probability density function for a non-discrete random variable on the basis of a set of finite original computational data points \cite{phonopy-phono3py-JPCM}. The lifetime values corresponding to each phonon frequency are thereby superimposed on the density distribution of phonon modes. An analysis of the phonon lifetimes, ie., how long a phonon mode lasts prior to decaying as a result of interactions is essential for the study of carrier relaxation, thermal/electrical transport and optoelectronic properties.
The lifetime of a particular phonon mode $\lambda$ ($\tau_\lambda$ is defined as the reciprocal of the phonon linewidth as $\tau_\lambda=1/[{2\Gamma_\lambda(\omega_\lambda)}]$ \cite{togo2015distributions}.
We note that phonon lifetimes can range between picoseconds (ps) to nanoseconds (ns) in prototypical semiconducting materials. Phonon lifetimes are primarily dominated by anisotropic phonon-phonon scattering and decreases with the increase in temperature due to increased scattering rates. We observe that, the phonon lifetimes are usually shorter for optical phonons compared to that of acoustic phonons.\\

From our calculations, we note that  in both Si and Ge, the lifetime of phonons decreases with increase in temperature. For Si, the phonon lifetimes range from $0-5$ ps for the Raman active modes around 15 THz, whereas the lifetimes are higher between $5-20$ ps for the lower frequency phonon modes. Similarly, for Ge, the phonon lifetimes range from $0-5$ ps for the Raman active modes, whereas the lifetimes for the lower frequency modes are slightly higher between $5-30$ ps. We note that the high phonon density heat map corresponds to the flat phonon dispersion between $12-15$ THz, which hosts the Raman active $E_{2g}$, $E_{1g}$ and $A_{1g}$ modes. In an effort to investigate the phonons, which contribute to majority of the larger lifetimes within the Brillouin zone (BZ), we have shown the phonon lifetimes from $M-K$ and $\Gamma-A$. It is observed that the higher concentration of phonon modes are localized around $12-15$ THz for 2-H Si, whereas higher phonon densities are found at around 8 THz for 2-H Ge, consistent with previous findings. Finally, the variation of the lifetimes for the full BZ mimics the contribution from $M-K$ for the case of hexagonal 2-H Si. This behavior is different for hexagonal 2-H Ge, where the lifetimes are about $30-50$ times less compared to that for the full BZ. Namely, for hexagonal Si, the lifetimes are maximum around 5 THz and decreases on either sides of the frequency spectrum, a feature different from the decreasing lifetimes from $M-K$.
For hexagonal Ge, the lifetimes are minimum around 5 THz and increases on either sides of the spectrum.\\ 




In order to address the anharmonicity of the phonon modes, we have performed calculations of the phonon linewidths as depicted in Fig. 7. The supercell method for evaluation of phonon anharmonicity is computationally quite intensive. Another less computationally demanding alternative, prevalent in literature,  pertains to the evaluation of the three-phonon scattering coefficients using the so-called 2n+1 theorem, implemented within DFPT. Here, we have utilized the quantum package D3Q, a tool for evaluation of the third order anharmonic force constants using DFPT and the 2n+1 theorem. The 2n+1 theorem, or Wigner's (2n+1) rule states that the (2n+1)th order perturbation correction to a non-degenerate energy can be obtained from the knowledge of the wavefunction up to the nth order. The above facilitates the estimation of the third derivative of total energy from the derivatives of wave functions and charge density in the ground state \cite{d3q}.\\

Figure 7 (a) and (b) below highlight the intrinsic phonon linewidth obtained for the hexagonal polytype of silicon and germanium at room temperature. The anharmonic three-phonon scattering involving the third order derivatives of the potential along with the scattering of three phonons with momentum \textbf{q}, \textbf{q$^\prime$}, \textbf{q$^{\prime\prime}$} provides an estimate for the linewidths in our formalism. The normal and umklapp scattering process are governed by the momentum conservation rules \textbf{q$^{\prime\prime}$} = $-$\textbf{q} $-$\textbf{q$^{\prime}$} and \textbf{q$^{\prime\prime}$} = \textbf{G} $-$\textbf{q} $-$\textbf{q$^{\prime}$} respectively. The linewidth broadening is expressed in units of $cm^{-1}$, and magnified by a factor of 10 for a demonstration of the relative orders of band and momentum resolved contributions. As expected, the three lowest acoustic phonon modes are observed to have smaller broadening compared to the higher frequency optical phonon modes.\\

Following our discussion of phonon lifetimes and linewidths, we would like to compare the results displayed in Figure 6 and 7. We note that, the phonon linewidths, are inversely proportional to that of the phonon lifetimes, as discussed in equation 6.
The lowest three acoustic modes  have the least linewidths, which broadens with the increase of phonon frequency. The above is consistent with the fact that the phonon lifetime decreases with the increase in the phonon frequency in Fig 6. Notice that, around the $\Gamma$ point of the Brillouin zone, the phonon linewidths increases till 300 $cm^{-1}$ (9 THz), followed by a decrease from $300-430$ $cm^{-1}$ (9-13 THz) and a subsequent increase beyond this frequency. Since the phonon dispersion is primarily dominated by the $\Gamma$ point, an opposite trend of decrease in lifetime, followed by an increase and subsequent decrease around 13 THz is observed in the calculated lifetimes for hexagonal Si as shown in Fig. 6. A similar decrease, increase and subsequent decrease in the phonon lifetimes were observed for 2-H germanium but at lower frequencies than silicon. Another interesting observation pertains to the fact, that the phonon lifetimes are much larger for in-plane directions compared to that of the out of plane direction in the Brillouin zone.
Figure 8 displays the temperature dependence of the phonon lifetimes for different frequencies as contour plots. We note that, in conformity with the previous discussion, the lifetime ranges from $5-100$ ps for both the lonsdaleite Si and Ge compounds. There is a steep decrease in the lifetimes as a function of temperature for the high frequency optical modes, whereas the lifetime decreases moderately for the mid frequency range and very slowly for the low frequency acoustic modes in both the polymorphs.
\begin{figure}[htbp]
    \centering
    \includegraphics[width=1\linewidth]{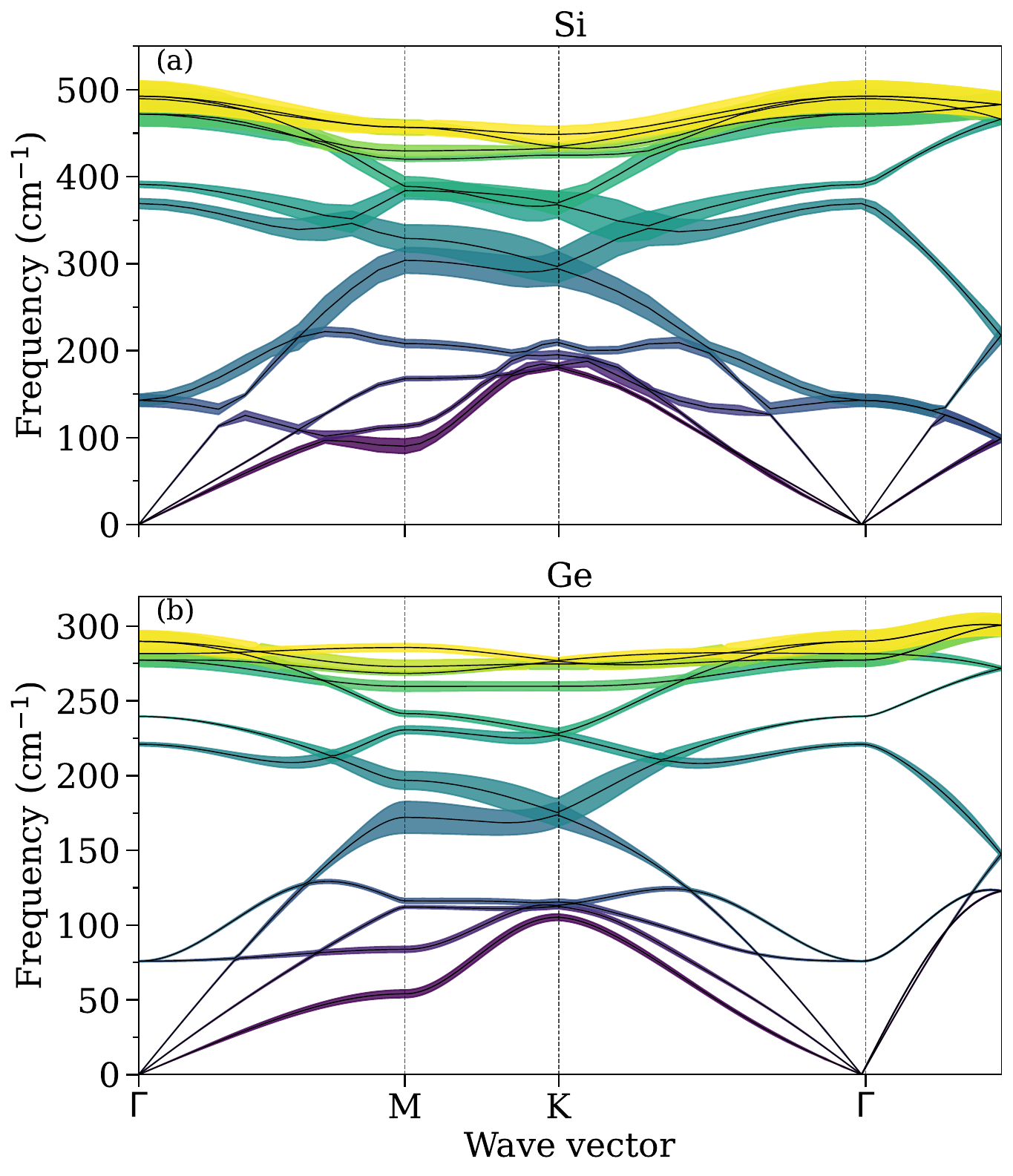}    
    \caption{Intrinsic phonon linewidth for hexagonal silicon (above) and hexagonal germanium (below) calculated at 300 K.}
    \label{fig:enter-label}
\end{figure}
\begin{figure}[htbp]
    \centering
    \includegraphics[width=1\linewidth]{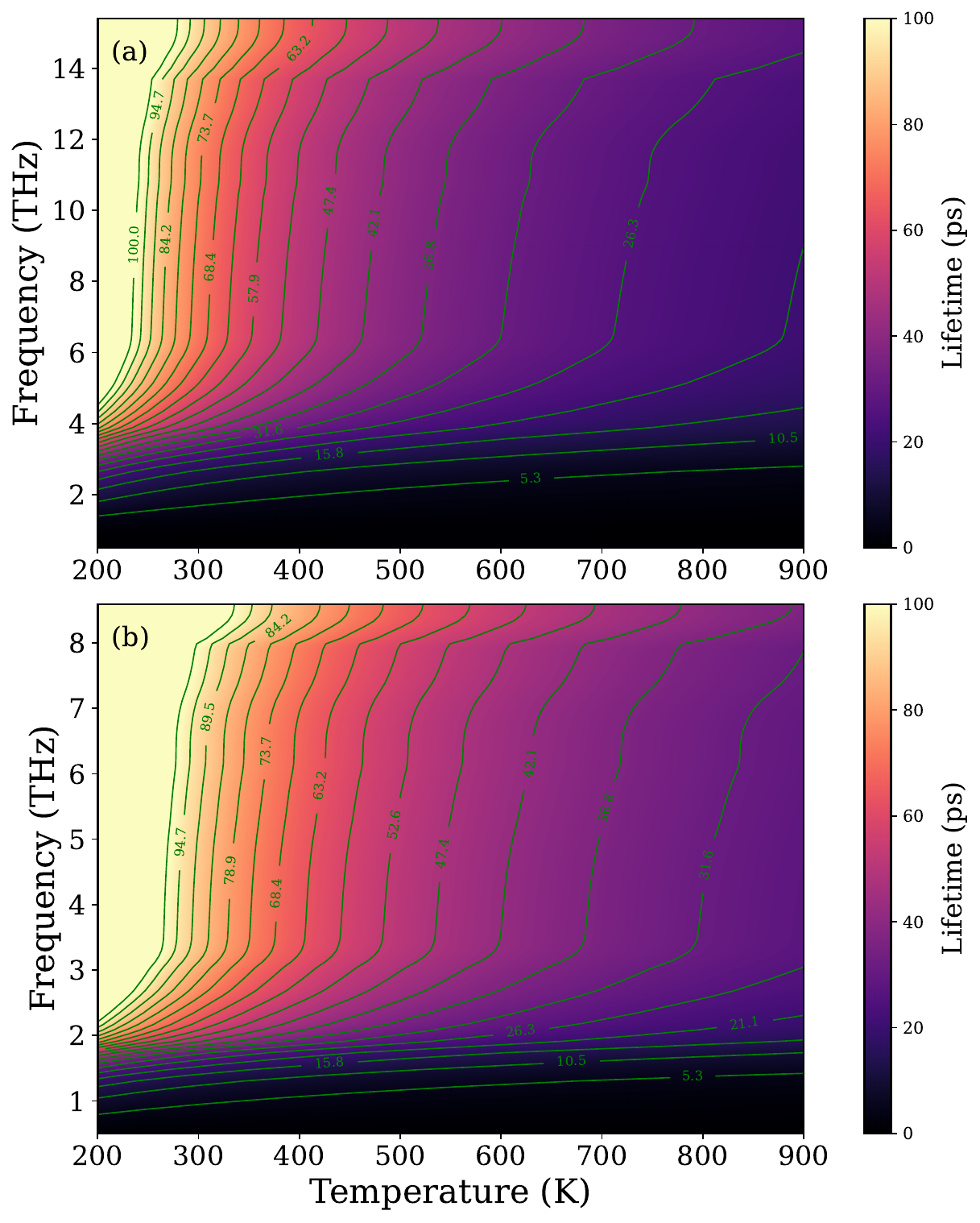}
    \caption{Temperature dependence of phonon lifetime for sH Si (a), and 2-H Ge (b).}
    \label{fig:placeholder}
\end{figure}
\subsection{Temperature dependent Mode Grüneisen Parameters and Anharmonicity}

An important parameter in the analysis of phonon stability, and anharmonicity is the so-called Grüneisen parameter. Primarily defined as the logarithmic pressure or volume derivative of the phonon frequencies, the Grüneisen parameter has been recently utilized as an efficient descriptor for the machine learning aided prediction of efficient thermoelectric materials and atomistic potential energy surfaces \cite{bhattacharjee2022thorough}, \cite{dutta2026machine}. Further, the singular behaviour of the Grüneisen parameter at a quantum critical point, due to the singularity of the thermal expansion has resulted in the parameter being identified as the foremost indicator of classical and quantum phase transitions \cite{soares2025universally}.
This parameter has recently surfaced as a tool to quantify caloric effect by variations of volume, magnetic field and interactions, and thereby found applications in ultracold quantum gases.\cite{squillante2023gruneisen},  Hence apart from it's paramount importance in anharmonic phonon dispersion relations, the parameter is recently being utilized in realization of quantum refrigeration \cite{yu2020gruneisen}.
It is customary to calculate the Grüneisen parameter utilizing Boltzmann-Gibbs staticstics using the following formula
\begin{equation}
\gamma = \left. \frac{\delta p}{\delta V} \right|_{V} = \frac{V \alpha_{V}B_{T}}{C_{V}} 
\end{equation}
where $\alpha_{V}$ is the coefficient of volumetric thermal expansion, $B_{T}$ is the isothermal bulk modulus and $C_{V}$ is the isochoric specific heat. Inelastic X-ray scattering and neutron diffraction are excellent experimental probes for materials characterization, which can be utilized to measure the Grüneisen parameter for real materials. In order to evaluate the  parameter from first principles calculations, one writes the thermodynamic Grüneisen parameter within the quasiharmonic approximation as
\begin{equation}
\gamma = \frac{\sum_{k} c_{V}^{k}\gamma_{k}}{C_{V}}
\end{equation}
where $c_{V}^{k}$ is the specific heat capacity of a harmonic oscillator with frequency $\omega_{k}$. The mode decomposed Grüneisen parameter can be expressed, in terms of the derivatives of the phonon frequncies with respect to changes of volume as
\begin{equation}
\gamma_{k} = -\frac{V}{\omega_{k}}\frac{\delta \omega_{k}}{\delta{V}}
\end{equation}

In this section of the manuscript, we have demonstrated the role of the Grüneisen parameter in quantifying anharmonicity for the Si and Ge polymorphs from first principles.
We have utilized the supercell approach for the calculations of phonon anharmonicity utilizing the Phonopy and Phono3py softwares for the evaluation of the mode decomposed Grüneisen parameters \cite{phonopy_phono3py}. Our phonon calculations performed at three volumes, for slightly different configurations ($\sim \pm 2 \%$), one configuration at the original volume at equilibrium, another one at unit cell volume $-2\%$ percent decrease and the final configuration at a volume of +2$\%$ increse of the original volume are done. In our electronic structure calculations, all the three configurations are relaxed under the fixed volume constraint.\\

The dynamical matrix elements and hence force constants are evaluated for each of the three different arrangements, which are then utilized for the computation of mode Grüneisen parameters, given by the following finite-difference equation:
\begin{equation}
\gamma(q\nu)=-\frac{V}{2[\omega(q\nu)]^2}<e(q\nu)|\frac{\Delta D(q)}{\Delta V}|e(q\nu) >
\end{equation}\\

In the above equation, $\omega(q\nu)$ is the frequency of the phonon at mode $\nu$ and the wave vector $q$, and the matrix element  of the change in the dynamical matrix $D$ with respect to volume $V$ is computed. Note that the expectation value of the above quantity is evaluated within the eigenvectors of the phonon polarization states, $e(q\nu)$. Hence, the computation of change in dynamical matrices require the construction of the increased, decreased and equilibrium unit-cell volumes, while the eigenfunctions determine the basis states for computing the expectation values of the matrices. Fig. 8 exhibits the mode decomposed Grüneisen parameters for the hexagonal Si and Ge compounds as a function of phonon frequency for room temperature (300 K). The different phonon modes are displayed with multiple colour palettes, each corresponding to a cluster of datasets originating from different k-points in the Brillouin zone from a particular phonon mode. We observe that the acoustic transverse phonon modes, across all k-points within the full Brillouin zone, are responsible for the wide range of values of the parameter, and is indicative of anharmonic phonon modes in the case of both 2-H silicon and 2-H germanium. We note that these frequencies range from $\sim$ $0-5$ THz for hexagonal Si and $\sim$ $0-3$ THz for hexagonal Ge respectively. Further, the deviations of the Grüneisen parameter is significantly less in germanium compared to that of silicon. The phonon modes in 2-H Si and 2-H Ge comprise primarily of three different parts: (i) the significantly scattered Grüneisen parameters with certain negative values in the low frequency phonon modes ($0-5$ THz for Si and $0-2$ THz for Ge), (ii) a slightly scattered set of Grüneisen parameters from $5-10$ THz for Si and $2-4$ THz for Ge) and finally (iii) almost no scattering of the parameters for the localized phonon modes with almost flat dispersion ($10-15$ THz for Si and $4-8$ THz for Ge respectively). \\

\begin{figure}[htbp]
    \centering
    \includegraphics[width=1\linewidth]{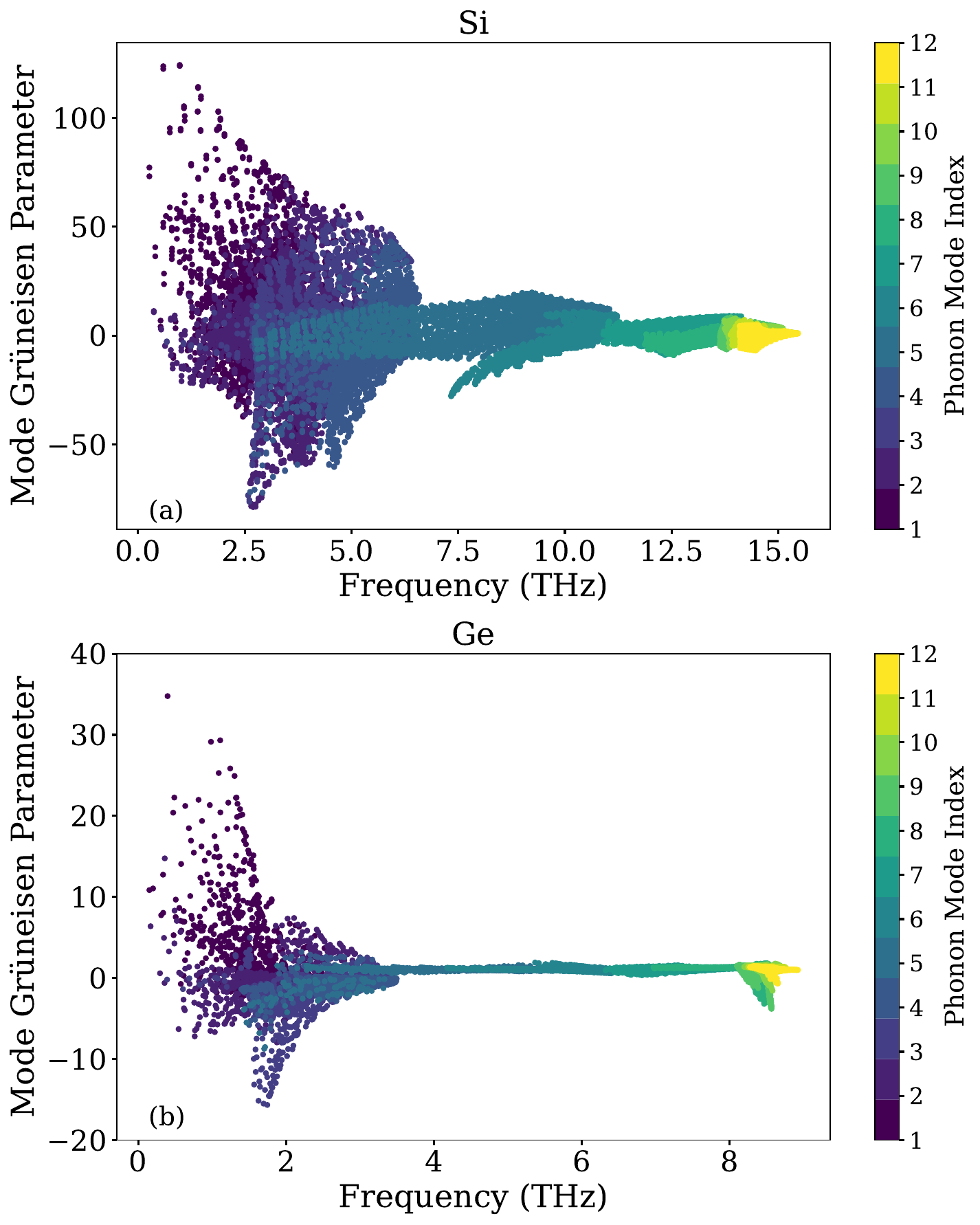}
    \caption{Mode Grüneisen parameter for 2-H Si (a) and 2-H Ge(b).}
    \label{fig:enter-label}
\end{figure}
\begin{figure}[htbp]
    \centering
    \includegraphics[width=1\linewidth]{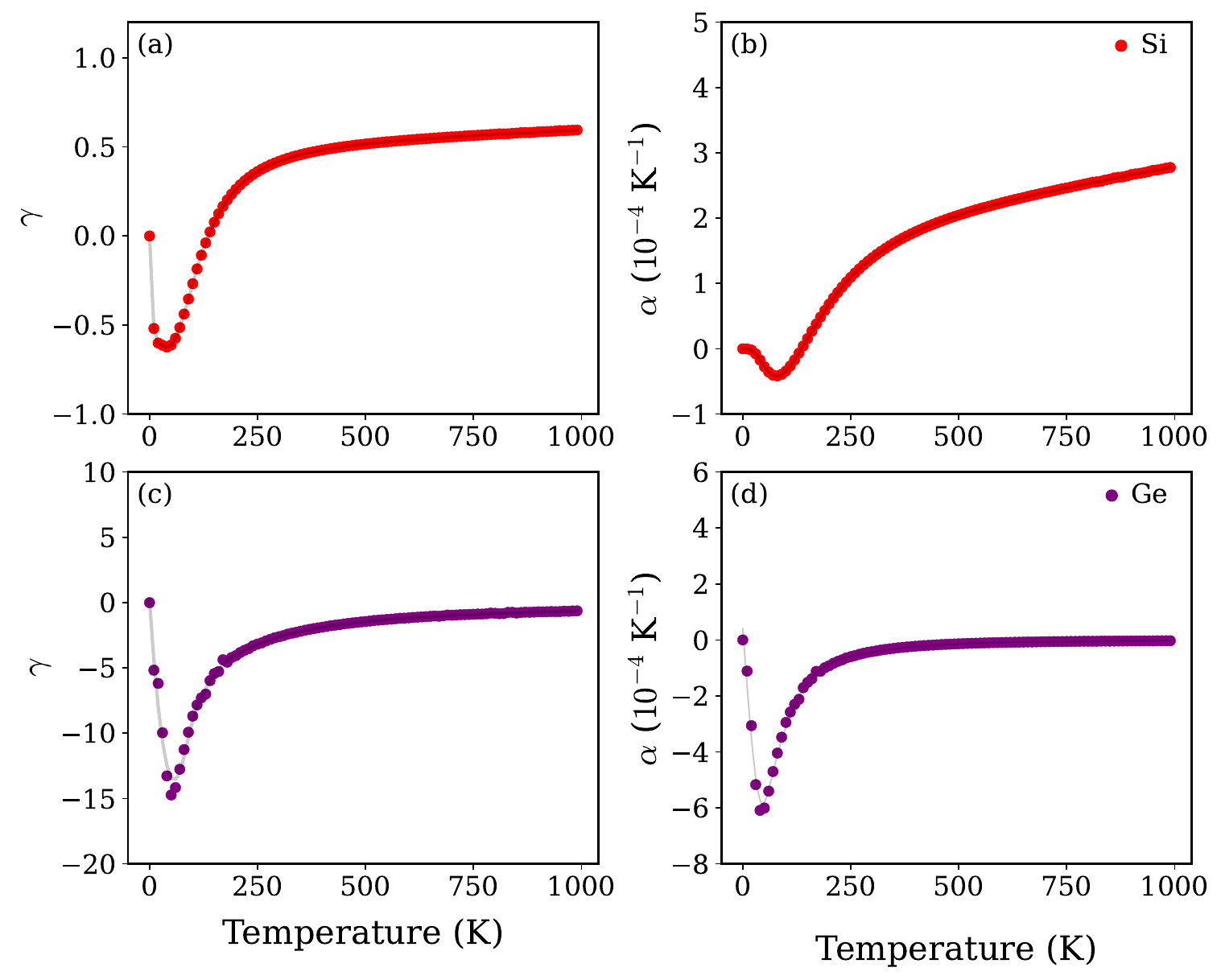}
    \caption{The temperature dependence of Grüneisen parameter ($\gamma$) and thermal expansion coefficient ($\alpha$) for 2-H Si (figures (a) and (b)) and 2-H Ge (figures (c) and (d)).}
    \label{fig:enter-label}
\end{figure}
The phonon modes are represented by color coding in the plots, the acoustic mode of lowest frequency being purple, and highest frequency optical mode being yellow. The dots represent each $q$ point for a particular phonon mode. It is evident from the figures that for both hexagonal silicon and germanium, the acoustic modes display a larger variation in the value of the Grüneisen parameter, whereas the optical modes of higher frequencies do not have any significant variation. For the case of hexagonal silicon, the acoustic phonons exhibit mode Grüneisen parameter varying from -75 to 125 units and optical phonons range in the region of $\pm25$ units whereas for germanium, the variation is from -25 to 35 units and $\pm10$ units respectively. In comparison, cubic silicon is a special case, where the acoustic modes have an average negative value for the Grüneisen parameter\cite{romero2020abinit}. There are three outlier points corresponding to the lowest frequency acoustic mode resulting in Grüneisen parameters in the range 60 to 180 units for hexagonal germanium, which have not been shown in the figure. It is also observed that for germanium, the highest frequency optical phonon modes (modes 11 and 12) exhibit slightly negative values of the parameter. This could be indicative of anharmonic contributions arising from phonon-phonon interactions or hint towards a possible phase transition in the material. \\

The temperature dependence of the thermal expansion coefficient and thermodynamic Grüneisen parameters were evaluated for a range of values $200-900$ K and depicted in Figure 10. 2-H Si displays a monotonous increase from approximately 2.7 to around 0.98 in the span of $200-900$ K. The curve saturates at higher temperature following a steep rise between $200-400$ K. This trend points to the increasing contributions from anharmonic phonon-phonon interactions with increase in temperature, consistent with the thermal population of higher-frequency phonon modes. 2-H Ge exhibits a drastic difference in the magnitude of the Grüneisen parameter $\gamma$ in comparison with silicon, evolving from approximately -3.8 at 200 K to about -1.2 at 900 K. The negative values suggest that certain phonon branches in this material exhibit anomalous volume dependence that could be associated with soft transverse acoustic or low-frequency optical modes, indicative of a possible phase transition which could arise from bond distortions. The gradual increase in the parameter (toward less negative values) with increasing temperature implies a reduction in the magnitude of anharmonic effects as higher-energy modes become thermally activated. 

\subsection{Transport and Thermal properties}
\begin{figure}[h]
    \centering
    \includegraphics[width=1\linewidth]{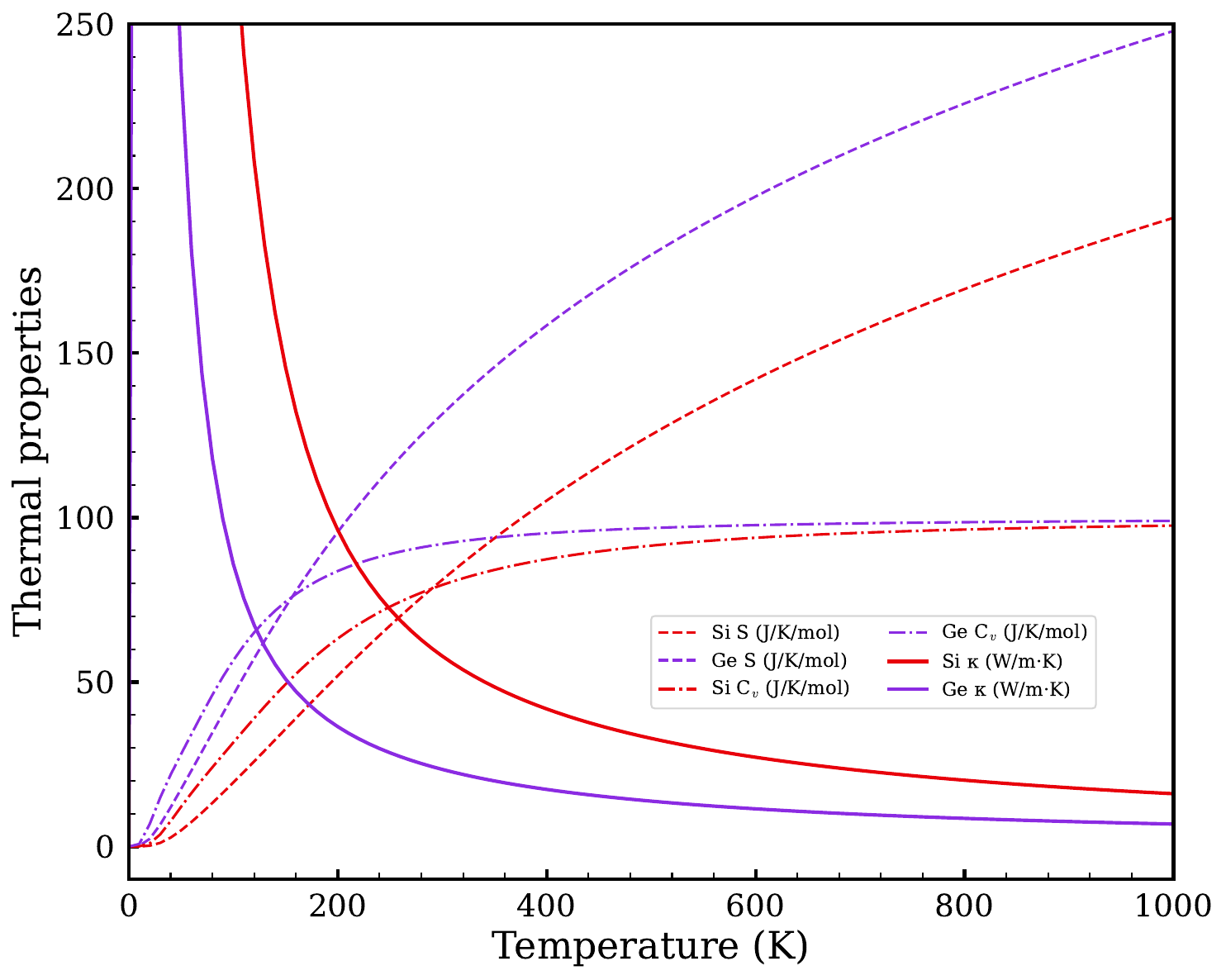}
    \caption{Thermal conductivity, entropy and specific heat at constant volume for a temperature range of 0 to 100 K for 2-H Si (red) and 2-H Ge (violet).}
    \label{fig:enter-label}
\end{figure}

Fig. 11 exhibits a comparison of the thermal conductivity, entropy and specific heat capacity for the lonsdaleite Si and Ge polymorphs. As depicted in the figure, the qualitative features of the plots for the thermal conductivity, entropy and specific heat follow a similar trend for both the hexagonal Si and Ge compounds. The plot for specific heat capacity is  increases as a polynomial function of temperature primarily due to phononic contributions and saturates for higher temperatures, mimicking the characteristics of similar potential thermoelectric materials.  We observe that the value of specific heat capacity $C_v$ undergoes a polynomial increase  up-till around 600 K, after which it tends to a saturate to a constant value of 100 J/K/mol, whereas for hexagonal Ge, the  saturation occurs above $\sim$ 400 K. In this case, the effect of thermal expansion on the specific heat at constant volume is not included, which might result in slight deviation of the values of specific heat $C_v$ obtained from  experimental results, particularly at higher temperatures. 
In our consideration, we have primarily focused on the effect of phonon modes on the thermal properties, since the phonons are the primary quasiparticles responsible for heat transport in these materials\cite{phonopy_phono3py}. \\

The lattice thermal conductivity is another important property to be explored following our investigation of the electronic structure and phonon calculations. According to the relaxation time approximation (RTA), when a system is perturbed by the application of heat, the mechanism by which equilibrium of the system is restored can be described by just the relaxation time $\tau$ of the system. Furthermore, application of the single-mode relaxation time approximation (SMRTA), results in each phonon mode relaxing independently to attain equilibrium, in other words, there is no coupling between the individual phonon modes, and when one mode is perturbed, the rest of the system tends to stay in equilibrium. In our formalism, the solution of the Peierls-Boltzmann equation has been incorporated for the above approximation to obtain the expression for lattice thermal conductivity \cite{phonopy_phono3py}. In Fig. 11, we have depicted the isotropic average of the thermal conductivity tensor.\\ 

From the figure, we note that at room temperature, hexagonal germanium exhibits a lower thermal conductivity of 23 W/m.K compared to that of hexagonal silicon, which is 59 W/m.K. The thermal conductivity tends to decrease with increase in temperature because of the higher phonon scatterings introduced at higher temperatures. In a later work, we plan to study the effect of alloys, interfaces and strain on the thermal conductivity of the Si and Ge allotropes, and tune the above parameters for improved thermoelectric efficiency.\\

 The incorporation of phonon-phonon and three-phonon interactions, associated scattering mechanisms for phonons,  phonon linewidths and lifetimes, have led to the prediction of finite lifetimes as suggested by our calculations. The phonon lifetimes have been quantified by evaluating the imaginary part of the phonon self-energy and can results in the broadening of the phonon spectral distribution. The corresponding phonon mean free path values hint at the scattering mechanisms involving transfer of momentum by phonons and the broadening of phonon lifetimes in real materials. In Figure 12, we have delineated the distribution of mean free paths along side the phonon density of states for hexagonal silicon and germanium respectively. Note that Fig. 12 in conjecture with Fig. 7 provides a detailed description of the frequency domains for the phonon mediated scattering mechanisms. From our first principles calculations, an increase in the phonon broadening towards higher frequency modes indicate increased phonon assisted scattering rates, which is confirmed by the depiction of the waterfall plot for the phonon mean free paths in these materials. We observe that higher phonon scattering rates are proportional to that of lower thermal conductivity values. For both the materials, Si and Ge, it is evident that the low frequency acoustic phonons correspond to lower scattering rates and hence higher mean free path values.\\
 \begin{figure}[h]
     \centering
     \includegraphics[width=1\linewidth]{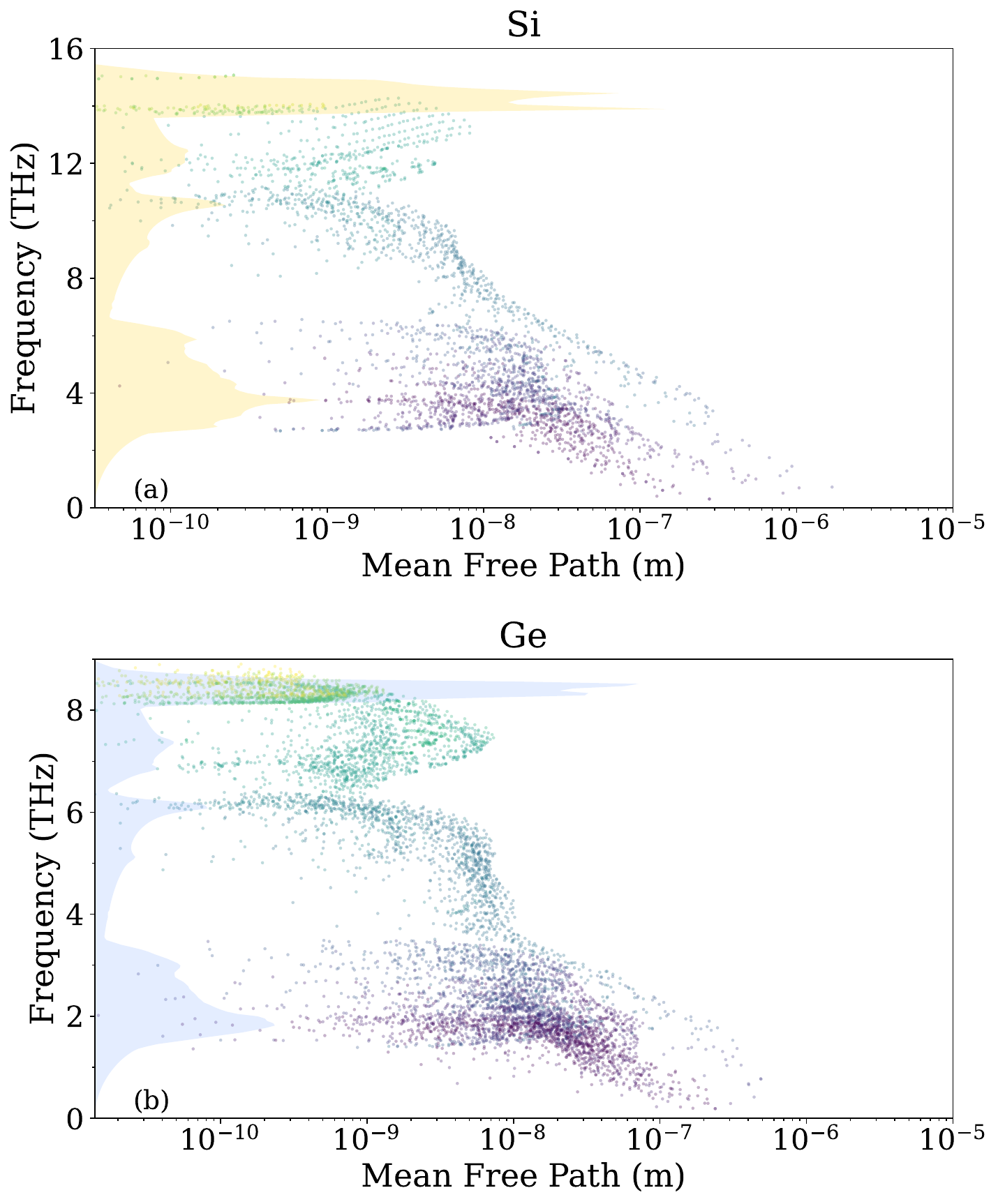}
     \caption{Variation of phonon mean free path with dos for hexagonal silicon and germanium.}
     \label{fig:waterfall}
 \end{figure}
 The phonon mean free path (MFP) is an important microscopic parameter that governs the lattice thermal conductivity. This quantity specifies the average distance covered by a phonon mode between successive scatterings. The mean free path of the phonon modes are obtained by performing a product of the phonon group velocities with that of the phonon relaxation times. Hence, we can express the phonon mean free path as
 \begin{equation}
     \lambda_\nu=v_\nu\tau_\nu
 \end{equation}
 where $\Lambda_\nu$, $v_\nu$ and $\tau_\nu$ are the MFP, phonon group velocity and relaxation time respectively \cite{phonopy_phono3py}.
 The figure 12 displays the waterfall representation generated using the thermoparser utility package and indicates the phonon mean free path for individual phonon branches at each wave vector along the Brillouin zone, thereby providing mode and frequency resolved insight into thermal transport mechanisms \cite{thermoparser_Spooner2024}. For both hexagonal silicon and germanium, a significant reduction in the phonon mean free path is observed in the range of  frequencies with associated high phonon density of states, indicating enhanced scattering due to higher availability of phonons for phonon–phonon interaction pathways. We further observe that the optical phonon modes exhibit shorter mean free paths compared to acoustic modes due to their stronger anharmonic scattering and comparatively smaller contribution to heat transport.\\

 \begin{table}[H]
\caption{\label{tab:table3}
Thermal and thermoelectric properties at 300 K}
\newcolumntype{C}[1]{>{\centering\arraybackslash}p{#1}}
\resizebox{\columnwidth}{!}{\begin{tabular}{|c|C{1.65cm}|C{1.75cm}|C{1.7cm}|C{1.7cm}|}
\hline
Polymorph&$\kappa$ W(mK)$^{-1}$&$\sigma /\tau_0$\newline ($\Omega$ms)$^{-1}$&S (V/K) \newline (at $E_f$)&C$_v$\newline J(K mol)$^{-1}$\\
\hline
2-H Si&57.9&3.37 $\times 10^{17}$&0.0003&79.57 \\
\hline
2-H Ge&22.9&$3.438 \times10^{18}$&$4.26 \times 10^{-5}$&92.13\\
\hline
\end{tabular}}
\end{table}

From an inspection of the orders of magnitude of the mean free paths in Fig. 12 (a), we infer that the pattern of distribution of mean free path primarily consists of three regions (i) for the low frequency region ($0-5$ THz for lonsdaleite Si and $0-3$ THz for lonsdaleite Ge), the mean free path is $\sim 10^{-7}$ m (ii) in the mid frequency regime ($5-10$ THz for hexagonal Si and $3-6$ THz for hexagonal Ge), the order of mean free path decreases to $\sim 10^{-8}$ m and (iii) in the neighborhood of the high frequency regime ($10-15$ THz for 2-H Si and $6-9$ THz for 2-H Ge respectively), the mean free path further drops by another order of magnitude to about $\sim 10^{-9}$ m. Hence, the mean free path for the Raman active optical modes at higher frequencies are two orders of magnitude lower than that of the anharmonic low frequency transverse acoustic modes in this material. A closer inspection of the mean free paths for propotypical thermoelectric and other technology relevant materials reveals that it typically ranges from $10^{-6} - 10^{-10}$ m \cite{sun2022}. Interestingly, the distribution of the mean free path follows a phase diagram similar to that of amorphous silicon \cite{fabian1997thermal}. In other words, both hexagonal Si and Ge possess three distinct areas (i) low frequency region ($0-20$ meV for hexagonal Si and $0-12$ meV for hexagonal Ge) dominated by acoustic vibrations of phonons moving ballistically with highest mean free paths, and least scattering rates (ii) mid frequency region ($0-40$ meV for 2-H Si and $12-25$ meV for 2-H Ge) primarily consisting of optical phonon vibrations propagating diffusively with lower mean free paths and higher scattering rates (iii) high frequency region ($\geq$ 40 meV for 2-H Si and $\geq$ 25 meV for 2-H Ge respectively ), where the electrons are essentially highly localized with the least mean free path.\\

Next, we investigate the phonon density of states and the corresponding mean free paths by comparison of the different regions in Fig. 12. We note that anharmonicity is mostly brought about by contributions from the low frequency transverse acoustic modes. However, the phonon density of states exhibits significant contribution from the higher frequency localized optical phonons. In order to quantify anharmonicity in the material, we have performed the following integral  $\frac{\int \delta \omega \: \omega DOS}{\int \delta \omega \, DOS}$. This parameter, referred to as phonon band center \cite{mukherjee2026phonon} is a descriptor that, together with the Grüneisen parameter indicates signatures of anharmonicity in a potential thermoelectric material. Previous efforts to quantify anharmonicity from phonon dispersion calculations, revealed an inverse relation between the Grüneisen parameter and the phonon band center, with values of phonon band center at low frequencies along with  large spread in the Grüneisen parameter to coexist in systems with predominantly anharmonic phonon modes \cite{mukherjee2026phonon}. Our calculation of the phonon band center reveals the centroid of frequency to be located at 9.75 THz for 2-H Si and 5.6 THz for 2-H Ge. Thus, even though the scattered range of Grüneisen parameter for the transverse acoustic modes suggest possible anharmonicity, the higher values of the phonon band center, biases the possibility of anharmonicity in these systems towards an inconclusive viability. Thus, we infer  based on our computational results that anharmonicity of phonon modes, if present in these systems, would contribute to, at best,  only a small fractional change in the energetics and quantities reliant on their derivatives.\\

\section{Conclusions}

In this manuscript, we have presented a detailed and systematic investigation of the electronic, vibrational, and transport properties of the hexagonal lonsdaleite silicon and germanium compounds using first-principles density functional theory and density functional perturbation theory. Compared to their cubic counterparts, the hexagonal polytypes exhibit distinct electronic and vibrational features which makes them potential material candidates with improved performance for thermoelectric and optoelectronic applications.
Our calculations for hexagonal silicon utilizing the popular GGA functional underestimates the indirect band gap to be 0.419 eV, whereas both hybrid functionals and GW approximation predict overestimated values of 0.984 eV and 0.95 eV. Hexagonal germanium, on the other hand, known to be a direct band gap semiconductor, is predicted to be a metallic system with no gap utilizing the GGA functional. We employed a recent yet successful exchange-correlation scheme, the meta-GGA SCAN functional for the  prediction of band structures for both the materials. The band gaps obtained for silicon and germanium are 0.769 eV and 0.202 eV respectively. The band gap for Ge is similar to that of the GW method (0.23 eV) whereas hybrid functionals overestimate the gap (0.283 eV). Our comparison of the difference in the charge density from GGA and SCAN functional reveals significant improvement in localization of the electronic density utilizing the meta-GGA formalism.\\

The computation of the phonon dispersion relation for both systems confirm lattice stability and lack of soft phonon modes, which is confirmed by symmetry considerations. The Raman mode decomposed phonon dispersion relations reveal anharmonic acoustic phonons dominate the  low frequency regime and are Raman inactive, while the localized optical phonon spectra at higher frequencies actively take part in Raman scattering. More specifically, the $A_{1g}, E_{1g}$ and $E_{2g}$ modes are Raman active in both the Si and Ge polymorphs. A detailed analysis of the Raman spectrum harnessing light sources, both unpolarized and circularly polarized, reveals the helicity-conserving and helicity-changing Raman modes in these materials. We observe that the Raman peaks correspond to 496 cm$^{-1}$ (A$_{1g}$, E$_{1g}$), 468 cm$^{-1}$ (E$_{2g}$) for 2-H Si, and 276 cm$^{-1}$ (A$_{1g}$, E$_{1g}$), 261 cm$^{-1}$ (E$_{2g}$) for 2-H Ge respectively. Interestingly, our calculations lead to the conclusion that the $E_{2g}$ mode changes the helicity of the circularly polarized incident light, whereas the $E_{1g}$ and $A_{1g}$ modes are helicity-conserving in nature.\\

The investigation of the phonon lifetimes, linewidths and mean free paths are instrumental for determining the role of anharmonicity, and phonon-phonon scattering rates in the transport and thermal properties of quantum materials. In this work, we have delineated the lifetimes of phonons as a function of phonon frequency, for different commensurate k-paths in the full Brillouin zone. Our study indicates a decrease in the lifetimes with increase in the phonon frequency, and a higher density of localized optical phonon modes at higher frequencies, responsible for helicity-dependent Raman scattering. Further, the BZ averaged phonon lifetimes range from $0-30$ ps for both of the lonsdaleite materials. For 2-H Si, the majority of the full BZ averaged lifetimes mimic the contributions from the M-K path of the Brillouin zone. The linewidths of the phonon modes evaluated to incorporate three-phonon scattering rates correlates inversely with the phonon lifetimes evaluated utilizing different first principles softwares. The temperature dependence of the phonon lifetimes exhibits three distinct regimes with the sharpest change in the phonon lifetimes vs temperature for the high frequency Raman active optical phonon modes.\\

In order to quantify the effects of anharmonicity, the frequency and temperature dependence of the mode Grüneisen parameter was computed for both 2-H Si and 2-H Ge compounds. The frequency dependent Grüneisen parameter reflects a widely scattered values for low frequency transverse acoustic phonons indicative of the possibility of anharmonicity. We also observed some negative values of Grüneisen parameter which hints towards significant warping of the bond environment indicating the possibility of a phase transition. The temperature dependent thermal expansion and Grüneisen parameters  are consistent with previously obtained results, although calculation of the phonon band center results in the role of anharmonicity in the materials to remain ambiguous. Finally, we dwelve into the waterfall plots for quantifying the effect of phonon-phonon scattering and role of phonon density of states in shaping the distribution of mean free paths for the two material systems. We note that both systems have three distinct domains with mean free paths varying over two orders of magnitude; the low frequency domain with ballistic phonon transport alongwith minimal phonon scattering, mid frequency domain with diffusive phonon transport and high frequency localized phonons with the least mean free paths. \\

The analysis of phonon modes highlight strong phonon–phonon scattering, particularly for optical phonon modes, resulting in reduced thermal conductivity relative to the cubic phases. The thermal conductivity has been observed to decrease with increasing temperature, with hexagonal germanium showing lower values than silicon, suggesting its suitability for thermoelectric applications where reduced heat transport is desirable. Thermoelectric transport calculations for hexagonal silicon reveal that optimal power factor occurs at carrier concentrations on the order of $10^{20}$ cm$^{-3}$ for hexagonal silicon. The combined picture indicates that that moderate band gaps, direct electronic transitions (for 2-H Ge), low thermal conductivity, and tunable thermoelectric coefficients render hexagonal silicon and germanium prospective candidate materials for next-generation electronic and energy conversion technologies. \\

Our goal in the near future, is the systematic exploration of the possibility of tuning the 2-H Si and 2-H Ge compounds as a function of composition of alloys of the two materials, for improved thermoelectric and optoelectronic applications. Further, we would like to investigate the effect of interfacing hexagonal 2-H Si and hexagonal 2-H Ge with ideal lattice matched candidates such as GaP and GaAs on the Raman active modes, mode Grüneisen parameters and mean free paths in these materials. Finally, the effect of converting the indirect to direct gap in hexagonal Si by the application of uniaxial strain, leading to efficient photoluminescence, improved LED devices and photovoltaics, enhanced phonon lifetimes and thermoelectric properties will be explored in greater depth for comparison with experimental findings.



%
%






\section{Acknowledgement}
The authors would like to thank Prof. Akshay Singh and Prof. Igor Mazin for valuable discussions, essential suggestions, and constructive feedback. We would also like to acknowledge the Sharanga high performance computing resources and research facilities at Birla Institute of Technology and Science Hyderabad, India for carrying out the project.\\ 

\section{Data Availability}

The data presented in the manuscript are available from
the following URL - https://github.com/Predictive-Quantum-
Simulations-Group/.

\section{Funding}

The corresponding author Prof. Suvadip Das would like to acknowledge the Early Career Grant - New Faculty Seed Grant (NFSG) for Phonons, Electron-phonon coupling and Superconductivity from First Principles endowed to him by the Birla Institute of Technology and Science Hyderabad.

\section{Author Contribution}

S.D. contributed towards conceptualisation, coordination, supervision, methodology and investigation of the project. S.D. wrote the original draft and edited the paper. L.S.M contributed to-
wards investigation, data curation, formal analysis, visualisation
and writing the original draft. U.A. contributed towards validation, visualisation and formal analysis. A.J. and S.S. contributed
towards data curation, and validation of the results.

\section{Conflict of Interest}

The authors would like to declare no potential conflicts of interest, either financial or intellectual, with any other research personnel or group during the dissemination of this work. 

\bibliographystyle{plain}
\bibliography{reference}
\end{document}